# Reducibility Among Fractional Stability Problems

Shiva Kintali [*]    Laura J. Poplawski [†][∥]    Rajmohan Rajaraman [‡][∥]

Ravi Sundaram [§][∥]    Shang-Hua Teng [¶]

## Abstract

"As has often been the case with NP-completeness proofs, PPAD-completeness proofs will be eventually refined to cover simpler and more realistic looking classes of games. And then researchers will strive to identify even simpler classes." –Papadimitriou (chapter 2 of [37])

In a landmark paper [39], Papadimitriou introduced a number of syntactic subclasses of **TFNP** based on proof styles that (unlike **TFNP**) admit complete problems. A recent series of results [12, 19, 6, 7, 8, 9] has shown that finding Nash equilibria is complete for **PPAD**, a particularly notable subclass of **TFNP**. A major goal of this work is to expand the universe of known **PPAD**-complete problems. We resolve the computational complexity of a number of outstanding open problems with practical applications.

Here is the list of problems we show to be **PPAD**-complete, along with the domains of practical significance: Fractional Stable Paths Problem (FSPP) [21] - Internet routing; Core of Balanced Games [41] - Economics and Game theory; Scarf's Lemma [41] - Combinatorics; Hypergraph Matching [1]- Social Choice and Preference Systems; Fractional Bounded Budget Connection Games (FBBC) [30] - Social networks; and Strong Fractional Kernel [2]- Graph Theory. In fact, we show that no fully polynomial-time approximation schemes exist (unless **PPAD** is in **FP**).

This paper is entirely a series of reductions that build in nontrivial ways on the framework established in previous work. In the course of deriving these reductions, we created two new concepts - preference games and personalized equilibria. The entire set of new reductions can be presented as a lattice with the above problems sandwiched between preference games (at the "easy" end) and personalized equilibria (at the "hard" end). Our completeness results extend to natural approximate versions of most of these problems. On a technical note, we wish to highlight our novel "continuous-to-discrete" reduction from exact personalized equilibria to approximate personalized equilibria using a linear program augmented with an exponential number of "min" constraints of a specific form. In addition to enhancing our repertoire of **PPAD**-complete problems, we expect the concepts and techniques in this paper to find future use in algorithmic game theory.

---

[*]College of Computing, Georgia Institute of Technology, kintali@cc.gatech.edu

[†]ljp@ccs.neu.edu

[‡]rraj@ccs.neu.edu

[§]koods@ccs.neu.edu

[¶]Microsoft Research New England and Boston University, steng@cs.bu.edu

[∥]College of Computer and Information Science, Northeastern University

# 1  Introduction

Intuitively, the notion of stability implies the absence of oscillations over time and encompasses the concepts of fixed points and equilibria. Stability is important in a variety of fields ranging from the practical - the Internet - to the theoretical - combinatorics and game theory. For important practical systems (e.g. Internet), the existence and computational feasibility of stable operating modes is of profound real-world significance. On the more abstract front, the study of stable solutions to combinatorial problems has a distinguished tradition dating back to, at least, the Gale-Shapley algorithm [17]. It is often the case, as with Nash's celebrated theorem [36], that *fractional* stable points are guaranteed to exist even when integral points don't. In this paper, we focus on fractional stability and resolve the computational complexity of a set of eight problems with applications to a variety of different domains. Six of these are pre-existing problems. Below we provide elaborate motivation for two of the pre-existing problems - **Fractional Stable Paths Problem (FSPP)** and **Core of Balanced Games**. The remaining four are: **Scarf's lemma**, a fundamental result in combinatorics with several applications [41], **Fractional Hypergraph Matching** [1], useful for modeling preferences in social-choice and economic systems, **FBBC**, the fractional version of the Bounded Budget Connection (BBC) game [30], which models decentralized overlay network creation and social networks, and **Strong Fractional Kernel** [2], of relevance to structural graph theory. In addition, we define two new concepts — personalized equilibria for matrix games and preference games — which are not only useful tools for carrying out reductions but also of independent interest.

**Fractional Stable Paths Problem.** Griffin, Shepherd and Wilfong [20] showed how BGP (Border Gateway Protocol, the routing mechanism of the Internet) can be viewed as a distributed mechanism for solving the Stable Paths Problem (SPP). They showed that there exist SPP instances with no integral stable solutions, a phenomenon that would explain why oscillation has been observed in Internet routes. Route oscillation is viewed as a negative, since it imposes higher system overheads, reorders packets, and creates difficulties for tracing and debugging. Subsequently, Haxell and Wilfong [21] introduced FSPP: a natural fractional relaxation of SPP with the property that a (fractional) stable solution always exists. Intuitively, FSPP can be viewed as a game played between Autonomous Systems that each assign fractional capacities to the different paths leading to a destination in such a way that they maximize their utility without violating the capacity constraints of downstream nodes. Understanding the computational feasibility of finding the equilibria of this game could help to develop techniques for stable routing in the Internet.

**Core of balanced games.** The notion of *core* in cooperative games is analogous to that of Nash equilibrium in non-cooperative games. Informally, a core is the set of all outcomes in which no coalition of players has an incentive to secede and obtain a better payoff, either viewed as a set (transferable utilities) or individually (non-transferable utilities). Necessary and sufficient conditions for the nonemptiness of the core in games with transferable utilities is given by the classic Bondareva-Shapley theorem [5, 43], which also yields a polynomial-time algorithm for finding an element in a nonempty core. Subsequently, in a celebrated paper, Scarf [41] generalized their result, developed certain sufficient *balance* conditions for the nonemptiness of the core in games with non-transferable utilities, and presented an algorithm for finding a point in the core. As noted by Jain and Mahdian in Chapter 15 of [37], "However, the worst case running time of this algorithm (like the Lemke-Howson algorithm) is exponential." Resolving the computational feasibility of finding the core in balanced games is of considerable significance in the theory of cooperative games.

**Personalized equilibria for matrix games - a generalization.** Imagine a business man-ufacturing and selling outfits consisting of a pant (solid or striped) and a shirt (cotton or wool). The manager of the location producing pants decides on the ratio of striped pants produced to solid pants while the manager at the location producing shirts decides on the ratio of cotton shirts produced to wool shirts. Each manager is then given half the total number of shirts and pants (in the proportions decided) and has to match them into outfits and sell them at her own location in such a way as to maximize her individual profits. Personalized equilibria for matrix games capture exactly this situation: each player chooses a distribution over her own actions, but then each player independently customizes the matching of her own actions to the actions of other players in such a way as to maximize individual payoff. The concept of personalized equilibria for matrix games generalizes a number of games and problems, including FSPP and FBBC.

**Preference games - a specialization.** Consider a world of bloggers where each blogger has a choice of actions. They can fill their blogs with original content or they can copy from the original content on others' blogs. Naturally, each blogger has a preference order over the content of the different bloggers (as well as their own). Also, of course, more cannot be copied from another blog than the amount that other blogger has written. The preference game models each blogger's choice of what percentage of his blog is original and what percentages are copied from which other blogs. Such preference games arise whenever each player has a preference among her actions, and her distribution over her actions is constrained by others' distributions. The definition of a preference game is surprisingly simple, making this a great candidate problem for reductions. In fact, preference games are reducible in polynomial-time to all the problems considered in this paper.

## 1.1 Our Contributions

Hewing to the dictum that a picture is worth a thousand words, we present a diagram (Figure 2) showing the different reductions. The takeaway is that all the eight problems of interest are **PPAD**-complete. To be precise, we show that for all these problems, the exact versions are in **PPAD**, and our reductions extend to natural approximation versions to show that there are no fully polynomial-time approximation schemes (unless **PPAD** is in **FP**). Our reductions build on prior work in intricate and involved fashion.

From a conceptual standpoint, we believe there is merit in the definitions of preference games and personalized equilibria. Preference games are very simple to describe and model a number of real-world situations, such as the blogger example mentioned earlier. Yet we can show that the set of equilibria of preference games can be nonconvex and in fact, are hard even to approximate. As a counterbalance we show that finding equilibria in the subclass of symmetric (for a natural notion of "symmetric") preference games is in **FP**. Personalized equilibria of matrix games are, we believe, a fascinating solution concept worthy of independent study. Not only do they model real-world situations as motivated earlier by the example of the apparel company, but they also constitute a natural generalization of a variety of predefined games, such as FSPP and FBBC. Our results on the hardness of approximating personalized equilibria for $k$-player games apply for $k \geq 4$. We show that finding personalized equilibria of 2-player games is in **FP**. The $k = 3$ case is open.

From a technical standpoint, we particularly wish to highlight our reduction from finding exact personalized equilibria to finding approximate personalized equilibria. To capture exact personalized equilibria, we write a linear program plus an exponential number of single-variable $min$ constraints. These are constraints specifying that the minimum of a subset of variables is 0. Using this specification, we prove the existence of rational equilibria. Furthermore, we reduce to approximate personalized equilibria by showing that an $\epsilon$-approximate equilibrium for sufficiently small $\epsilon$ points us to a subset of the variables that can be set to 0 to simultaneously satisfy all of the *min* constraints, leaving us with a polynomially-sized feasible linear program for an exact equilibrium. With this reduction in hand and an additional technical bound on the size of short feasible vectors, we are then able to carry through the reduction to End-of-the-Line to show that personalized equilibria is in **PPAD**.

## 1.2 Related Work

Nash profoundly changed game theory by demonstrating the existence of mixed equilibria [35, 36]. Decades later, on the computational front [37], the complexity class **TFNP** was introduced by Megiddo and Papadimitriou [34]. Papadimitriou's seminal work [39] not only defined a number of syntactic subclasses of **TFNP** (including **PPAD**), but also proved that a variety of problems, including discrete versions of Brouwer's fixed point theorem and Sperner's lemma, are **PPAD**-complete. The problem of finding Nash equilibria was left open. Recently, a series of papers comprising different author combinations of the two teams, Daskalaikis-Goldberg-Papadimitriou [12, 19] and Chen-Deng-Teng [6, 7, 8, 9] culminated in establishing that *approximating* Nash equilibria with two players, 2-NASH, is hard. The reductions in our work build on the framework established in these papers.

BGP has been the focus of much attention since its inception [40, 45]. As mentioned earlier SPP was introduced by Griffin, Shepherd and Wilfong [20] to explain the nonconvergence of BGP [47]. Haxell and Wilfong [21] defined FSPP and proved the existence of an equilibrium using Scarf's lemma and a compactness-type argument. They left open the complexity of finding an equilibrium. Our reduction from personalized equilibria to End-of-the-Line is a different approach that generalizes the Haxell-Wilfong existence result while preserving computational tractability. Kintali [27] presented a distributed algorithm for finding an $\epsilon$-approximation for FSPP that is guaranteed to converge, although no bounds are given on the time-to-convergence (our results imply a polynomial time bound is unlikely).

Cooperative games, the study of mechanisms to sustain and enforce cooperation among willing agents, has a rich and extensive literature [10, 18, 11, 16, 29]. As mentioned earlier, in a celebrated paper Scarf [41] generalized the classical Bondareva-Shapley theorem [5, 43] result and developed an algorithm for finding a point in the core of balanced games with non-transferable utilities. More recently, Markakis and Saberi [33], Immorlica, Jain and Mahdian [23] studied certain classes of games with non-transferable utilities in the context of the Internet; however, it is unclear that their problems are even in **TFNP**. Scarf's paper [41] also contains Scarf's lemma, an important result in combinatorics which played a crucial part in the FSPP existence proof of Haxell and Wilfong [21]. Aharoni and Holzman [2] proved that every clique-acyclic digraph has a strong fractional kernel, and Aharoni and Fleiner [1] proved that every hypergraphic preference system has a fractional stable matching. Both of these proofs are based on Scarf's lemma. The computational complexity of these problems was left unresolved.

The BBC game, introduced in [30, 31], builds on a large body of work in network formation games [24, 4]. A direct precursor to BBC games was introduced by Fabrikant et al. [15]. Fractional BBC games were introduced in [31], but the problem of finding an equilibrium was left open.

## 2 The Class PPAD

A major contribution of this paper is to expand the set of problems known to be **PPAD**-complete. The class **PPAD** (*Polynomial Parity Argument in a Directed graph*) was introduced by Papadimitriou in [39], which defined a number of syntactic classes in the semantic class **TFNP**, or the set of all total search problems. A search problem $\mathcal{S}$ consists of a set of inputs $I_\mathcal{S} \subseteq \Sigma^*$ such that for each $x \in I_\mathcal{S}$ there is an associated set of solutions $\mathcal{S}_x \subseteq \Sigma^{|x|^k}$ for some integer $k$. For each $x \in I_\mathcal{S}$ and $y \in \Sigma^{|x|^k}$, it is decidable in polynomial time whether or not $y$ is in $\mathcal{S}_x$. A search problem is *total* if $\mathcal{S}_x \neq \emptyset$ for all $x \in I_\mathcal{S}$. **TFNP** is the set of all total search problems [34]. Since every member of **TFNP** is equipped with a mathematical proof that it belongs to **TFNP**, a number of syntactic classes can be defined based on their proof styles. The complexity class **PPAD** is the class of all search problems whose totality is proved using a directed parity argument.

Problems in **PPAD** are reducible to the END OF THE LINE problem. In END OF THE LINE, we are given a finite directed graph in which each node has at most one outgoing edge and at most one incoming edge. The input to the problem is not a complete list of the nodes and edges; such a list may be exponentially large in the size of the input. Instead, we are given an initial source node and a circuit. The circuit takes a node name as input and in polynomial time returns the *next* node (the other end of the outgoing edge from the input node) and the *previous* node (the other end of the incoming edge into the input node). If the input node is a source (or sink), null is returned as the previous (or next) node. The problem for END OF THE LINE is to find a sink or a source other than the initial source.

Throughout this paper, we use PROBLEM A $\leq_P$ PROBLEM B to mean "There exists a polynomial time reduction from finding a stable point in PROBLEM A to finding a stable point in problem PROBLEM B."

## 3 Preference Games

In this section, we define a very simple game, the *preference game*. Each player has a preference list across the set of players and must assign weight to each player. No player may put more weight on another player than that player puts on itself. A best response for a player occurs when that player cannot move weight from a lower preference player to a higher preference player. We show in Section 3.2 that when preferences are symmetric, it is very easy to find an equilibrium in which all weights are either 0 or 1. However, in Section 3.3 we show that the set of equilibria in general preference games may not be convex, implying that we cannot hope to find an equilibrium using convex programming, and in Section 3.4, we show that finding an equilibrium in general preference games is **PPAD**-hard. In Section 3.5, we define an $\epsilon$-approximate equilibrium for the preference game and extend our **PPAD**-hardness result to approximate equilibria. Our notion of approximation carries though all of the reductions in later sections, so we prove that there are no fully polynomial-time approximation schemes (unless **PPAD** is in **FP**) for computing stable points in any of the problems discussed in this paper. Finally, in section 3.6, we define the *degree* of a preference game, and show that any preference game can be reduced to a preference game with constant degree.

### 3.1 Preference Games

In a preference game with a set $S$ of players, each player's strategy set is $S$. Each player $i \in S$ has a preference relation $\succeq_i$ among the strategies.[1] For strategies $j$ and $k$, $j \succeq_i k$ indicates that player $i$ prefers $j$ at least as much as $k$. When it is clear from context that we are talking about the preferences for player $i$, we write $j \succeq k$ instead of $j \succeq_i k$. Each player $i$ chooses a *weight distribution*, which is an assignment $w_i : S \to [0, 1]$ satisfying two conditions: (a) the weights add up to 1: $\sum_{j \in S} w_i(j) = 1$; and (b) the weight placed by $i$ on $j$ is no more than the weight placed by $j$ on $j$: $w_i(j) \leq w_j(j)$ for all $i, j \in S$.

---

[1] A preference relation is a binary relation that is transitive and complete.

Given weight assignments $w_i$, $w_i'$, and $w_{-i}$ such that $(w_i, w_{-i})$ and $(w_i', w_{-i})$ are both feasible, we say $w_i$ is *lexicographically at least* $w_i'$ (with respect to $w_{-i}$) if for all $j \in S$, $\sum_{k \succeq_i j} w_i(k) \geq \sum_{k \succeq_i j} w_i'(k)$. We say that $w_i$ is *lexicographically maximal* (implied: with respect to $w_{-i}$) if $(w_i, w_{-i})$ is feasible and $w_i$ is lexicographically at least every assignment $w_i'$ such that $(w_i', w_{-i})$ is feasible. An equilibrium in a preference game is an assignment $w = \{w_i : i \in S\}$ such that $w_i$ is lexicographically maximal with respect to $w_{-i}$ for all $i \in S$.

Every preference game has an equilibrium, a fact which can be shown using standard fixed-point theorems; we defer the proof to Section 4, where we show the existence and PPAD-membership of a more general class of equilibria.

---

PREFERENCE GAME: Given a set of players $[n]$, each with strategy set $[n]$, and a preference relation $\succeq_i$ among the strategies for each player $i$. Find a feasible weight assignment $w$ such that for all $i$, $w_i$ is lexicographically maximal with respect to $w_{-i}$.

---

### 3.2 Symmetric Preference Games

In a *symmetric preference game*, the players are ordered $\{1, \ldots, n\}$. Given the order of the players, we have the following symmetry in the preferences: if $i \leq j$, and if $i \succeq_j j$, then $j \succeq_i i$. In other words, if a player $j$ is later in the order than $i$, and if $j$ prefers $i$ over itself, than the earlier player $i$ also prefers $j$ over itself.

**Theorem 3.1.** *In any symmetric preference game, an equilibrium in which all weights are $0$ or $1$ can be found in polynomial time.*

*Proof.* If our preference rules obey this style of symmetry, we can use Algorithm 1 to find an equilibrium.

---

**Algorithm 1** Finding an equilibrium in a symmetric preference game

1: Sort the players into their symmetry order.
2: Set all weights to $-1$.
3: **for** $i = 1 \ldots n$ **do**
4:   **if** $w_i(i) = -1$ **then**
5:     Assign $w_i(i) = 1$.
6:     **for** $j = i+1 \ldots n$ **do**
7:       Assign $w_i(j) = 0$.
8:       **if** $j \succeq_i i$ **then**
9:         Assign $w_j(j) = 0$.
10: **for** $i = 1 \ldots n$ **do**
11:   **if** $w_i(i) = 0$ **then**
12:     Find the player $j$ with $w_j(j) = 1$ that is highest in $i$'s preference list.
13:     Assign $w_i(j) = 1$.
14:     Assign $w_i(k) = 0$ for all other $k \neq j$.

---

Since each player has weight 1 assigned to exactly one strategy, Algorithm 1 assigns a feasible set of weights. To show that Algorithm 1 finds an equilibrium, we must show that the results of the algorithm obey the following. (a) If $w_i(i) = 1$, then there is no $j$ such that $j \succeq_i i$ with $w_j(j) = 1$, and (b) if $w_i(i) = 0$, then there is some $j$ such that $j \succeq_i i$ with $w_j(j) = 1$.

To show (a): consider the point in the algorithm at which $w_i(i)$ is set to 1. By this point, we have already looked through all $j$ ahead of $i$ in the ordering. Since $w_i(i)$ is still $-1$, for each $j$ for which we assigned $w_j(j) = 1$, none had $i \succeq_j j$. By symmetry, this means that no $j$ ahead of $i$ in the ordering has $w_j(j) = 1$ and $j \succeq_i i$. Now, for all $j$ following $i$ in the ordering, if $j \succeq_i i$, then we assign $w_j(j) = 0$ immediately after we assign $w_i(i) = 1$.

To show (b): Consider the point at which we assigned $w_i(i) = 0$. We had just assigned $w_j(j) = 1$ for some $j$ ahead of $i$ in the order. We found that $i \succeq_j j$, which by symmetry implies that $j \succeq_i i$, as required. □

### 3.3 Non-Convexity

Although symmetric preference games have a simple equilibrium which can be found in polynomial time, general preference games are more complex. In this section, we show that the set of equilibrium for a preference game may not be convex.

**Theorem 3.2.** *There exists an instance of the preference game for which the set of equilibria is not convex.*

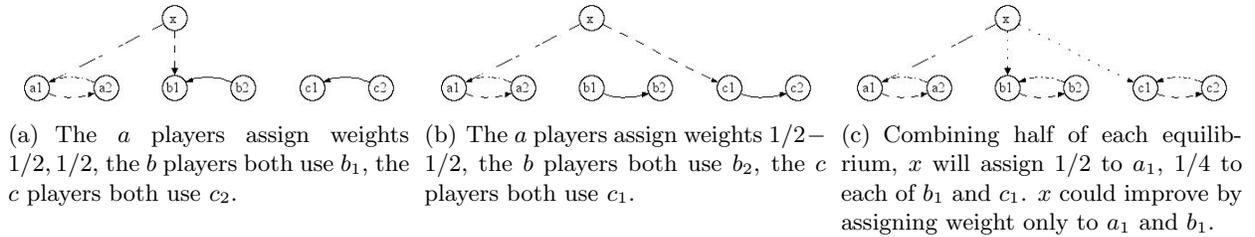

(a) The $a$ players assign weights $1/2, 1/2$, the $b$ players both use $b_1$, the $c$ players both use $c_2$.

(b) The $a$ players assign weights $1/2-1/2$, the $b$ players both use $b_2$, the $c$ players both use $c_1$.

(c) Combining half of each equilibrium, $x$ will assign $1/2$ to $a_1$, $1/4$ to each of $b_1$ and $c_1$. $x$ could improve by assigning weight only to $a_1$ and $b_1$.

Figure 1: Example of an instance of the preference game for which the equilibrium set is not convex.

*Proof.* Consider the following instance of the preference game. We have 3 sets of 2 players each, $a_1, a_2, b_1, b_2, c_1, c_2$, and one additional player, $x$. The preference lists for these nodes are: $a_1$: $(a_2, a_1)$; $a_2$: $(a_1, a_2)$; $b_1$: $(b_2, b_1)$; $b_2$: $(b_1, b_2)$; $c_1$: $(c_2, c_1)$; $c_2$: $(c_1, c_2)$; $x$: $(a_1, b_1, c_1, x)$. (Each list gives strategies in order from most preferred to least preferred.) We now show two equilibria whose linear combination is not an equilibrium. In equilibrium $w$ (figure 1(a)): $w_{a_1}(a_1) = \frac{1}{2}$, $w_{a_1}(a_2) = \frac{1}{2}$, $w_{a_2}(a_2) = \frac{1}{2}$, $w_{a_2}(a_1) = \frac{1}{2}$, $w_{b_1}(b_1) = 1$, $w_{b_2}(b_1) = 1$, $w_{c_1}(c_2) = 1$, $w_{c_2}(c_2) = 1$, $w_x(a_1) = \frac{1}{2}$, $w_x(b_1) = \frac{1}{2}$. In equilibrium $w'$ (figure 1(b)): $w'_{a_1}(a_1) = \frac{1}{2}$, $w'_{a_1}(a_2) = \frac{1}{2}$, $w'_{a_2}(a_2) = \frac{1}{2}$, $w'_{a_2}(a_1) = \frac{1}{2}$, $w'_{b_1}(b_2) = 1$, $w'_{b_2}(b_2) = 1$, $w'_{c_1}(c_1) = 1$, $w'_{c_2}(c_1) = 1$, $w'_x(a_1) = \frac{1}{2}$, $w'_x(c_1) = \frac{1}{2}$. It is easy to verify that $w$ and $w'$ are both equilibria, and in a solution $\lambda \cdot w + (1 - \lambda) \cdot w'$ (for any $\lambda > \frac{1}{4}$) (figure 1(c) shows $\lambda = \frac{1}{2}$), player $x$ would do better by moving more weight to its second preference. Therefore, the convex combination of $w$ and $w'$ is not an equilibrium. □

### 3.4 PPAD Hardness

We show that finding an equilibrium in preference games is **PPAD**-hard. We will follow the framework of [12], which shows that finding a Nash equilibrium in a degree-3 graphical game is **PPAD**-hard, using a reduction from the **PPAD**-complete problem 3-D BROUWER. In this problem, we are given a 3-D cube in which each dimension is broken down into $2^{-n}$ segments – thereby dividing the cube into $2^{3n}$ cubelets. We are also given a circuit that takes as input the 3 coordinates of the center of a cubelet (each as an $n$-bit number) and returns a 2-bit number that represents one of four 3-D vectors: either $(1, 0, 0)$, $(0, 1, 0)$, $(0, 0, 1)$, or $(-1, -1, -1)$. A solution to the 3-D BROUWER instance is a cubelet vertex such that the set of 8 results obtained by running the circuit on each of the 8 cubelets surrounding the vertex contains each of the four vectors at least once.

As in [12], we will construct a set of gadgets to simulate various arithmetic operators, logical operators, arithmetic comparisons and other operators. We then follow their framework to systematically combine these gadgets to simulate the input boolean circuit and to encode the geometric condition of discrete fixed points in the 3-D BROUWER instance. In the preference game we construct, we specify the preference relation of any player $P$ by an ordered list of a subset of the players, with the last element being $P$, also referred to as the "self" strategy. When we say that a player $P$ *plays itself* with weight $v$, we mean that $P$ assigns a weight of $v$ to strategy $P$. We'll engineer the payoffs such that the game is only in equilibrium if the weights assigned by certain players to *themselves* successfully echo the inputs and outputs of 8 copies of the circuit that surround a solution vertex of the 3-D BROUWER instance.

For this reduction, we require the following sets of players.

1. One player for each of the 3 dimensions (the *coordinate players*). If the graph is an equilibrium, each coordinate player plays itself with weight equal to its coordinate of the 3-D BROUWER solution vertex.

2. One player for each of the bits of each of the 3 coordinates (the *bit players*). In order to force these players to correctly represent the bits, we need some additional players. Assuming we've correctly calculated the first $i-1$ bits of coordinate $x$ (call them $x_0, \ldots, x_{i-1}$), we can create the $i^{th}$ bit as follows. One player will play itself with weight $p_i = x - \sum_{j=0}^{i-1} \frac{x_j}{2^j}$. The bit player will play itself with weight equal to the $i^{th}$ bit. If $p_i \geq \frac{1}{2^i}$, then this bit should be 1. Otherwise, it should be 0. Therefore, in order to properly extract the bits, we create the following four types of players.

   (a) HALF player: In any equilibrium in which a given player plays itself with weight $a$, the HALF player will play itself with weight $\frac{a}{2}$.

   (b) DIFF player: In any equilibrium in which two given players play themselves with weights $a$ and $b$, the DIFF player will play itself with weight $a - b$.

   (c) VALUE player: In any equilibrium, the VALUE player plays itself with weight $\frac{1}{2}$. This can be easily created by combining a player whose first preference is itself with a HALF player.

   (d) LESS player: In any equilibrium in which two given players play themselves with weights $a$ and $b$, respectively, the LESS player plays itself with weight 1 iff $a \geq b$, and plays itself with weight 0 otherwise. (Actually, the LESS player we create will be inaccurate if $a$ and $b$ are very close, which we discuss more below.)

3. One player simulates each type of gate used in the circuit of the 3-D BROUWER instance. For this, we create 3 more types of players.

(e) **AND player:** In any equilibrium in which two given players play themselves with weights $a$ and $b \in \{0, 1\}$, the AND player will play itself with weight $a \wedge b$.

(f) **OR player:** In any equilibrium in which two given players play themselves with weights $a$ and $b \in \{0, 1\}$, the OR player will play itself with weight $a \vee b$.

(g) **NOT player:** In any equilibrium in which a given player plays itself with weight $a \in \{0, 1\}$, the NOT player will play itself with weight $\neg a$.

4. Finally, we need to ensure that the graph is in equilibrium if and only if all four vectors are represented in the results of the 8 circuits. As in [12], we will represent the output of each circuit using 6-bits, one each for $+x, -x, +y, -y, +z, -z$. Now, the 4 possible result vectors are represented as 100000, 001000, 000010, and 010101. We can use these circuit results with only two additional types of players to feed back into the original coordinate players. First, we will create an OR player for each of the 6 bits (over the 8 vertices), which yields a result of six 1's if and only if this is a solution vertex. Therefore, an AND player for each coordinate will all return 1 if and only if this is a solution vertex; at least one of the coordinates will be 0 otherwise. We can turn this around using a NOT player for each coordinate, so that we get all 0's if and only if this is a solution vertex. Finally, we need the last two new player types, which we'll use to add these results back to a copy of the original coordinates (the result will be the original coordinate player).

(h) **COPY player:** In any equilibrium in which a given player plays itself with weight $a$, the COPY player will also play itself with weight $a$.

(i) **SUM player:** In any equilibrium in which two given players play themselves with weights $a$ and $b$, the SUM player will play itself with weight $\min(a + b, 1)$.

If the coordinates represented a solution vertex to the 3-D BROUWER instance, then all the values we've added back in will be zero; so the coordinate players cannot do better by changing their strategies. On the other hand, if the coordinates do not form a solution vertex, then at least one of the values is 1, so that the coordinate player will have incentive to change strategies and play more weight on itself.

We now describe how to create the new types of players (gadgets) required for the reduction. For each of these gadget definitions, we assume we are given a preference game such that in any equilibrium, player $X$ plays itself with weight $v_1$ and player $Y$ plays itself with weight $v_2$. For the first three gadgets, we assume $v_1, v_2 \in \{0, 1\}$. For the rest of the gadgets, we assume $v_1, v_2 \in [0, 1]$.

**OR**$(X, Y)$

We can add a new node $R = \text{OR}(X, Y)$ that will play itself with weight $v_1 \vee v_2$ in any equilibrium. Create a node $R_1$ with preference list $(X, Y, R_1)$. Let node $R$'s preference list be $(R_1, R)$. Now, if $v_1$ and/or $v_2$ is 1, then $R_1$ will play $R_1$ with weight 0, so $R$ will play itself with weight 1. If both $v_1$ and $v_2$ is 0, then $R_1$ will play itself with weight 1, so $R$ will play $R_1$ with weight 1 and $R$ with weight 0.

**NOT**($X$)

We can add a new node $N = \text{NOT}(X)$ that will play itself with weight $\neg v_1$ in any equilibrium. Let node $N$'s preference list be $(X, N)$. Clearly, $N$ will play $X$ as much as $v_1$ and will play $N$ with the remainder.

**AND**($X, Y$)

We can add a new node $A = \text{AND}(X, Y)$ that will play itself with weight $v_1 \wedge v_2$ in any equilibrium. Assemble the OR and NOT gadgets NOT(OR(NOT($X$), NOT($Y$))).

**SUM**($X, Y$)

We can add a new node $S = \text{SUM}(X, Y)$ that will play itself with weight $max(1, v_1 + v_2)$ in any equilibrium. Create a node $S_1$ with preference list $(X, Y, S_1)$. Let node $S$'s preference list be $(S_1, S)$. Now, clearly node $S_1$ will play $S_1$ with weight $max(0, 1 - v_1 - v_2)$, and node $S$ will play $S_1$ that same amount. So node $S$ will play itself with weight $1 - max(0, 1 - v_1 - v_2)$. In other words, if $v_1 + v_2 \geq 1$, then $S$ will play itself with weight 1. Otherwise, $S$ will play itself with weight $1 - 1 + v_1 + v_2 = v_1 + v_2$, as desired.

**DIFF**($X, Y$)

We can add a new node $D = \text{DIFF}(X, Y)$ that will play itself with weight $v_1 - v_2$ if $v_1 > v_2$, or 0 otherwise in any equilibrium. Create a node $D_1$ with preference list $(X, D_1)$. $D_1$ will play itself with weight $1 - v_1$. Now set the preference list for $D$ to $(D_1, Y, D)$. $D$ will play itself with weight $\min(0, 1 - (1 - v_1) - v_2) = \min(0, v_1 - v_2)$, as desired.

**COPY**($X$)

We can add a new node $C = \text{COPY}(X)$ that will play itself with weight $v_1$ in any equilibrium. Create a node $C_1$ with preference list $(X, C_1)$. $C_1$ will play itself with weight $1 - v_1$. Set the preference list for node $C$ to $(C_1, C)$. $C$ will play $C_1$ with weight $1 - v_1$, leaving weight $v_1$ on $C$.

**DOUBLE**($X$)

We can add a new node $M = \text{DOUBLE}(X)$ that will play itself with weight $\min(1, v_1 * 2)$ in any equilibrium. Create player $M_1 = \text{COPY}(X)$ and set $M$ as $\text{SUM}(X, M_1)$.

**LESS**($X, Y$)

Given $\epsilon_l$ ($0 < \epsilon_l \leq \frac{1}{2}$), We can add a new node $L = \text{LESS}(X, Y)$ to the game that in any equilibrium will play only itself if $v_1 - v_2 \geq \epsilon_l$, and will play $L_1$ (for a new node $L_1$) if $v_1 \leq v_2$. First create $D = \text{DIFF}(X, Y)$. Then create $M_1 = \text{DOUBLE}(D)$. For $i = 1$ to $-\log \epsilon_l$, create player $M_{i+1} = \text{DOUBLE}(M_i)$. Call the last DOUBLE player node $L$ and the extra player for the sum player of the last DOUBLE player node $L_1$. If $v_1 \leq v_2$, the DIFF player will return 0, so player $L$ will play the result of multiplying 0 by 2 many times, or 0. If $v_1 - v_2 \geq \epsilon_l$, player $L$ will play the max of 1 and $(v_1 - v_2) * 2^{-\log \epsilon_l} = (v_1 - v_2) * \frac{1}{\epsilon_l} \geq \frac{\epsilon_l}{\epsilon_l} = 1$.

**HALF**($X$)

We can add a new node $H = \text{HALF}(X)$ that will play itself with weight $v_1/2$ in any equilibrium. Create a node $H_1$ with preference list $(X, H_1)$. $H_1$ will play itself with weight $1 - v_1$. Then create two more nodes: $H_2$ and $H_3$. Node $H_2$ has preference list $(H_1, H_3, H_2)$. Node $H_3$ has preference list $(H_1, H, H_3)$. Set the preference list for node $H$ to be $(H_1, H_2, H)$. Each of $H$, $H_2$, and $H_3$ will use its first choice with weight $1-v_1$, leaving $v_1$ for its other two choices. Then, we have $w_H(H) + w_H(H_2) = v_1$, $w_{H_2}(H_2) + w_{H_2}(H_3) = v_1$, and $w_{H_3}(H_3) + w_{H_3}(H) = v_1$. In any equilibrium, it must be true that $w_H(H_2) = w_{H_2}(H_2)$, $w_{H_2}(H_3) = w_{H_3}(H_3)$, and $w_{H_3}(H) = w_H(H)$. Solving this gives $w_H(H) = w_H(H_2) = w_{H_2}(H_2) = w_{H_2}(H_3) = w_{H_3}(H_3) = w_{H_3}(H) = \frac{v_1}{2}$.

As in [12], our LESS player plays the specified action (itself, in our case) with weight 1 if $v_1 \geq v_2 + \epsilon_l$, and plays itself with weight 0 if $v_1 \leq v_2$, but will play some unspecified fraction on itself if $v_2 < v_1 < v_2 + \epsilon_l$. We use the LESS player to extract the bits representing the coordinates of a cubelet to be passed into the circuit. This procedure is identical to that of [12]. Let $X$ denote the x-coordinate player, and let $X_1 = \text{COPY}(X)$. For $i$ from 1 through $n$, we create players $B_i = \text{LESS}(2^{-i}, X_i)$ and $X_{i+1} = \text{DIFF}(X_i, \text{HALF}^i(B_i))$, where $\text{HALF}^i$ indicates applying the HALF gadget $i$ times. It can be shown that as long as $x$ is not too close to a multiple of $2^{-n}$, we will extract its $n$ bits correctly. If this is not the case, however, we will not properly extract the bits, and our circuit simulation may return an arbitrary value. We resolve this problem using the same technique as in [12]: we compute the circuit for a large constant number of points surrounding the vertex and take the average of the resulting vectors. Since these details are almost identical to that of [12, Lemma 4], we omit them.

Based on the above gadgets and the framework from [12], we get the following.

**Theorem 3.3.** 3-D BROUWER $\leq_P$ PREFERENCE GAME. □

### 3.5 Approximate equilibria

Given the hardness of finding exact equilibria in preference games, a natural next question is whether it is easier to find approximate equilibria. We define an $\epsilon$-equilibrium of a $k$-player preference game to be a set of weight distributions $w_1, \ldots, w_k$ that satisfy the following conditions for every player $i$: (a) $\sum_j w_i(j) = 1$; (b) for each $j$, $w_i(j) \leq w_j(j) + \epsilon$; and (c) for each $j$, either $\sum_{\ell:\ell \succeq j} w_i(\ell) \geq 1 - \epsilon$ or $|w_i(j) - w_j(j)| \leq \epsilon$. The problem of finding an $\epsilon$-equilibrium is $\epsilon$-APPROXIMATE PREFERENCE GAME.

**Theorem 3.4.** BROUWER $\leq_P$ $\epsilon$-APPROXIMATE PREFERENCE GAME. *Thus, it is* **PPAD**-*hard to find an $\epsilon$-equilibrium for preference games for $\epsilon$ inverse polynomial in $n$.*

*Proof.* Our proof follows the framework of [8, 9] for proving the hardness of approximating Nash equilibria in 2-player games. This framework starts with a high-dimensional discrete fixed point problem, BROUWER, which is also **PPAD**-complete. The input to BROUWER is a Boolean circuit that assigns a color from $\{1, ..., n, n+1\}$ to each interior node of an $n$-dimensional grid $\{0, 1, ..., 8\}^n$. This grid has about $2^{3n}$ cells, each of which is an $n$-dimensional hypercube. The discrete fixed point is defined to be a panchromatic simplex inside a hypercube. This framework of [8, 9] uses a new geometric condition for discrete fixed points, which requires that the average of $n^3$ sampled points in the interior of the targeted panchromatic simplex is inverse-polynomially close to the zero vector. The rest of the proof follows the framework of [12].

Our broad definition of an $\epsilon$-equilibrium poses additional technical challenges which did not occur in the reductions of [8, 9]. In particular, in the presence of errors, our Boolean gadgets only approximately simulate the Boolean operations, while in previous reductions, the Boolean gadgets are precise. We prevent magnification of errors in the Boolean simulation by strategically adding a LESS gadget to correct errors after each logic step.

We focus on bounding the errors for the gadgets of Theorem 3.3 and the addition of the extra LESS gadgets. Other details closely match those of [8, 9, 12].

Let $\epsilon_l$ (the measure of the fragility of our LESS gadget) be a real number such that $\epsilon \leq \epsilon_l^3$. Then, we have the following error bounds.

**Lemma 3.5.** *Assuming node $X$ plays itself with weight $v_1'$, $v_1 - \epsilon_l \leq v_1' \leq v_1 + \epsilon_l$, and node $Y$ plays itself with weight $v_2'$, $v_2 - \epsilon_l \leq v_2' \leq v_2 + \epsilon_l$, each of the boolean gadgets plays itself within $\pm(2\epsilon_l + 6\epsilon)$ of the correct value for the correct $v_1$ and $v_2$ inputs.*

*Proof.* **OR**

If $v_1$ and/or $v_2$ is 1, then $v_1'$ and/or $v_2'$ is at least $1 - \epsilon_l$, and node $R_1$ will play $R_1$ with weight at most $\epsilon_l + \epsilon$, so $R$ will play $R$ with weight at least $1 - \epsilon_l - 2\epsilon$. If both $v_1$ and $v_2$ are 0, then $v_1'$ and $v_2'$ are at most $\epsilon_l$, and node $R_1$ will play $R_1$ with weight at least $1 - 2\epsilon_l - 2\epsilon$, so $R$ will play $R$ with weight at most $2\epsilon_l + 3\epsilon$.

**NOT**

If $v_1 = 1$, $v_1'$ is at least $1 - \epsilon_l$, and node $N$ will play itself with weight at most $\epsilon_l + \epsilon$. If $v_1 = 0$, $v_1'$ is at most $\epsilon_l$, and node $N$ will play $N$ with weight at least $1 - \epsilon_l - \epsilon$.

**AND**

The AND gadget concatenates other new players to get $\neg(\neg v_1 \vee \neg v_2)$. Each NOT may add at most one additional $\epsilon$ error to the given value, and the OR may add up to $3\epsilon$ error (on top of the sum of the errors from both inputs). So the AND player will return a value within an additive $2\epsilon_l + 6\epsilon$ of the correct 0 or 1 answer. □

**Lemma 3.6.** *Each of the arithmetic gadgets plays itself within $\pm 5\epsilon$ of the correct value for the input it is given.*

*Proof.* **SUM**

Node $S_1$ will play $S_1$ with weight $w(S_1T) \in [max(0, 1 - v_1' - v_2' - 2\epsilon), max(0, 1 - v_1' - v_2' + 2\epsilon)]$. So node $S$ will play $S$ with weight $w_S(S) \in [v_1' + v_2' - 3\epsilon, v_1' + v_2' + 3\epsilon]$, unless $w_{S_1}(S_1) = 0$, which means $v_1' + v_2' \geq 1 - 2\epsilon$. In this case, node $S$ will play $S$ with weight at least $1 - \epsilon$.

**DIFF**

Node $D_1$ will play $D_1T$ with weight $w_{D_1}(D_1) \in max(0, 1 - v_1' - \epsilon), max(0, 1 - v_1' + \epsilon)]$. Node $D$ will play $D$ with weight $w_D(D) \in [max(0, v_1' - v_2' - 3\epsilon), max(0, v_1' - v_2' + 3\epsilon)]$, unless $w_{D_1}(D_1) = 0$ which means $v_1' \geq 1 - \epsilon$. In this case, node $D$ will play $D$ with weight at least $1 - v_2' - 2\epsilon$ and at most $1 - v_2' + \epsilon$ (not $2\epsilon$ because we cannot underfill the strategy with weight 0).

**COPY**

Node $C_1$ will play $C_1$ with weight at least $1 - v_1' - \epsilon$ and at most $1 - v_1' + \epsilon$. Node $C$ will play $C$ with weight at least $v_1' - 2\epsilon$ and at most $v_1' + 2\epsilon$.

**HALF**

Node $H_1$ will play $H_1$ with weight $w_{H_1}(H_1) \in [1 - v'_1 - \epsilon, 1 - v'_1 + \epsilon]$, and each other player will play its second and third preferences with total weight between $1 - w_{H_1}(H_1) - \epsilon$ and $1 - w_{H_1}(H_1) + \epsilon$. Each other player will play itself half of this amount plus or minus $3\epsilon$ (this is easy to verify by writing the system of inequalities and checking the extreme points). Therefore, node $H$ plays $H$ with weight at least $\frac{v'_1}{2} - 4\epsilon$ and at most $\frac{v'_1}{2} + 4\epsilon$.

**DOUBLE**

The DOUBLE gadget consists of a copy player, which adds at most $2\epsilon$ error, and a sum player, which adds at most $3\epsilon$ error on top of the sum of the errors in the two inputs. Therefore, node $M$ plays $M$ with weight at least $2v'_1 - 5\epsilon$ and at most $2v'_1 + 5\epsilon$. □

**Lemma 3.7.** *The LESS player will play itself with weight $< \epsilon_l$ if it is given $v'_1, v'_2$ such that $v'_1 \leq v'_2$, and with weight $> 1 - \epsilon_l$ if $v'_1 - v'_2 \geq \epsilon_l$.*

*Proof.* **LESS**

The LESS gadget inherits its susceptibility to error from its initial DIFF player (which was, in the exact equilibrium case, non-zero if and only if $v_1 < v_2$). For the case where $v_1 < v_2$, we can account for the errors of the DOUBLE players (used to repeatedly amplify the difference) simply by adding extra iterations of DOUBLE. Since we stipulated that $\epsilon \leq \epsilon_l^3$, a value that started $\leq 5\epsilon$ will remain $< \epsilon_l$, even after doubling enough times to push a value $\geq \epsilon_l$ to a value over 1 (including extra multiplications to account for the DOUBLE errors). Therefore, the LESS player will play itself with weight less than $\epsilon_l$ if $v'_1 \leq v'_2$ and with weight greater than $1 - \epsilon_l$ if $v'_1 - v'_2 \geq \epsilon_l$. □

**Lemma 3.8.** *By using a LESS gadget after each boolean logic gadget, we can ensure that the output from each gate is at most $\epsilon_l$ away from the correct output.*

*Proof.* After a single gate (if the inputs are within additive $\epsilon_l$ of the correct 0 or 1 inputs), a player will play itself at least $1 - 2\epsilon_l - 6\epsilon$ if the correct answer is 1, and at most $2\epsilon_l + 6\epsilon$ if the correct answer is 0 (based on the analysis in the proof of Lemma 3.5). Call this player OUTPUT and the value it plays itself $v$. Then, we only need to add a player CONSTANT-HALF who plays itself with weight close to $\frac{1}{2}$, and a LESS player, CORRECTION = LESS(OUTPUT, CONSTANT-HALF).

CONSTANT-HALF can be made up of a player who plays itself with weight at least $1 - \epsilon$ and at most 1 (its first preference is for itself) and a HALF player, who by Lemma 3.6 will play itself with weight at least $\frac{1-\epsilon}{2} - 5\epsilon$ and at most $\frac{1}{2} + 5\epsilon$.

We know that if the correct answer was 0, then $v \leq 2\epsilon_l + 6\epsilon < \frac{1-\epsilon}{2} - 5\epsilon$, so CORRECTION will play itself with weight $< \epsilon_l$ (by Lemma 3.7), and if the correct answer was 1, then $v \geq 1 - 2\epsilon_l - 6\epsilon > \frac{1}{2} + 5\epsilon + \epsilon_l$, so CORRECTION will play itself with weight $> 1 - \epsilon_l$ (again by Lemma 3.7). □

After the corrections, we're left with the following possible errors due to the $\epsilon$-approximation. We have small errors in the bit extraction, which are no larger than the parallel errors in [12] (they verify that these small error values will not affect the final result). We also have small errors (at most $\epsilon_l$) coming out of the circuit. As in [8, 9], we will repeat the circuit a polynomial number of times and take the average in order to override any errors from the LESS gadgets in the bit extraction.

Taking an average of two results requires 3 steps: first we divide each "bit" in half (we cannot take the average of the entire values because we have a max value of 1 for any player, so the average of two 1's would come out to $\frac{1}{2}$). Here, we may pick up $4\epsilon$ of error for each of the two results. Then, we sum the two. The total error so far is at most $11\epsilon$. Finally, we take half of the sum, which also divides the error in half, but may add up to an additional $4\epsilon$ of error, for a total additional error of at most $9.5\epsilon$ from taking the average of 2 results.

We can add LESS gadgets periodically during the averaging and during the final OR, AND and NOT of the results to keep our total errors under $\epsilon_l$. In other words, if this is a solution vertex for BROUWER, then we will have 6 players, each playing at most $\epsilon_l$. If this is not a solution vertex, then at least one of the 6 players will play at least $1 - \epsilon_l$.

Suppose we have an $\epsilon$-equilibrium in this game, and the x-coordinate player is playing value $x$. This is a SUM player, and the extra player from the SUM gadget must be playing between $1 - x - \epsilon$ and $1 - x + \epsilon$. Therefore, the sum of the two values it is adding (a copy of the coordinate player and the feedback NOT player) must be between $x - 3\epsilon$ (if this player overfills each of its top strategies by $\epsilon$) and $x + 3\epsilon$ (if this player underfills each of its top strategies by $\epsilon$). We know that the copy player must be playing the same value as the coordinate player to within $2\epsilon$ (between $x - 2\epsilon$ and $x + 2\epsilon$). Adding this range to a number $\geq 1 - \epsilon_l$ cannot possibly give something in the range $[x - 3\epsilon, x + 3\epsilon]$, so the feedback player must be playing a value at most $\epsilon_l$ on itself (since we know the feedback player will play either a value $\leq \epsilon_l$ or a value $\geq 1 - \epsilon_l$), and the correct feedback must be 0, so this is a valid fixed point.

□

## 3.6 Constant degree preference games

For a given preference game, define in($v$) (resp., out($v$)) of a player $v$ to be the set $\{u : v \succ_u u\}$ (resp., $\{u : v \prec_v u\}$). We define the in-degree (resp., out-degree) of a player $v$ to be $|\text{in}(v)|$ (resp., $|\text{out}(v)|$). The degree of the player is defined to be the sum of the in-degree and the out-degree of the player. The in-degree (resp., out-degree, degree) of the preference game is defined to be the maximum, over all nodes, of the in-degree (resp., out-degree, degree) of the node. Notice that this is the same as the degree in a directed graph in which each player is represented by a node, and an edge from $u$ to $v$ means that $u$ prefers $v$ over itself. DEGREE $d$ PREFERENCE GAME is the problem of finding an equilibrium in a preference game with constant degree $d$.

Notice that the players defined in Section 3.4 all have out-degree at most 2. There is no implicit constant bound on the in-degree, but by adding COPY gadgets (which have out-degree 1) we can guarantee in-degree at most 2. Furthermore, since COPY gadgets have out-degree 1, we can make sure that the overall degree of the preference game is at most 3. This automatically implies that it is **PPAD**-hard to find an equilibrium even in a preference game with degree 3. We will use this fact in later sections, where we show **PPAD**-hardness of several other problems via reductions from constant degree preference games. In addition, we have the following reduction.

**Theorem 3.9.** PREFERENCE GAME $\leq_P$ DEGREE $d$ PREFERENCE GAME

*Proof.* Given a preference game over player set $[n] = \{1, \ldots, n\}$, with the sum of the lengths of the preference lists equal to $m$. Assume that each player exists in the preference list (ahead of "self") for at most $m'$ other players.

**Reducing to a preference game with constant out-degree (at most $c+1$), with $O(m+n)$ players.**

Suppose player $i$ in the original game has preference list $j_1, j_2, \ldots, j_k$. Let $d = \lceil \frac{k}{c} \rceil$. Create $2d$ new players, split into two sets: $I = \{i_1, \ldots, i_d\}, I' = \{i'_1, \ldots, i'_d\}$. For ease of notation, we will also refer to player $i_d$ as $i^*$, since this is the player that will play itself the same amount that the original player $i$ should play itself.

Set the preference list for the new player $i_1$ to $j_1^*, j_2^*, \ldots j_c^*, i_1$. For new player $i_h$ ($h > 1$), set the preference list to $i'_{h-1}, j^*_{(h-1)c+1}, j^*_{(h-1)c+2}, \ldots, j^*_{hc}, i_h$. For each new player $i'_h$ ($h \geq 1$), set the preference list to $i_h, i'_h$.

**Equilibrium in the original preference game maps to an equilibrium in the new preference game.**

The map will be as follows: Suppose we are given weights $w(i,j)$ for the original game, where $w(i,j)$ is the weight player $i$ puts on player $j$. We will set the weights $w^*$ in the new preference game as follows. Again, assume the preference list for player $i$ in the original game is $j_1, j_2, \ldots, j_k$.

- $w^*(i_h, j^*) = w(i,j)$ for all $j^*$ in the preference list of $i_h$.
- $w^*(i_h, i'_{h-1}) = \sum_{l=1}^{(h-1)c} w(i, j_l)$
- $w^*(i_h, i_h) = 1 - \sum_{l=1}^{hc} w(i, j_l)$
- $w^*(i'_h, i_h) = 1 - \sum_{l=1}^{hc} w(i, j_l)$
- $w^*(i'_h, i'_h) = \sum_{l=1}^{hc} w(i, j_l)$

Notice,

$$
\begin{aligned}
w^*(i^*, i^*) &= w^*(i_d, i_d) \text{ (by definition of } i^*) \\
&= 1 - \sum_{l=1}^{dc} w(i, j_l) \text{ (from map above)} \\
&= 1 - \sum_{l=1}^{\lceil \frac{k}{c} \rceil c} w(i, j_l) \text{ (by definition of } d) \\
&= 1 - \sum_{l=1}^{k} w(i, j_l) \text{ (we can ignore the } \lceil \rceil \text{ since the pref list stops after } k \text{ items)} \\
&= w(i,i)
\end{aligned}
$$

In order to verify that this is an equilibrium in the new game, we must check the following

1. $w^*(i,j) \leq w^*(j,j)$ for all $i,j$.
2. $w^*(i,i) + \sum_{j \neq i} w^*(i,j) = 1$ for all $i$.
3. If $w^*(i,j) > 0$, and if $i$ prefers $a$ over $j$, then $w^*(i,a) = w^*(a,a)$.

All three of these are trivial for players in $I'$, so we will verify the conditions for players in $I$. First consider condition 1 for each weight placed by a player in set $I$.

- $w^*(i_h, a^*) = 0$ unless $a^*$ is in the preference list for $i_h$. If $a^*$ is in the preference list, then $a^* = a'_p$ for some player $a$ from the original game with $p = \lceil$ length of $a$'s preference list $/c \rceil$, and $w^*(a^*, a^*) = w(a, a)$. By the map above, $w^*(i_h, a^*) = w(i, a)$. Since $w(i, a) \leq w(a, a)$, $w^*(i_h, a^*)$ obeys condition 1.

- $w^*(i_h, i'_{h-1}) = \sum_{l=1}^{(h-1)c} w(i, j_l)$. But we know from the map that $w^*(i'_{h-1}, i'_{h-1}) = \sum_{l=1}^{(h-1)c} w(i, j_l)$, so $w^*(i_h, i'_{h-1})$ also obeys condition 1.

Next, check condition 2 for each player in set $I$. The total weight placed by player $i_h$ is

$$w^*(i_h, i_h) + w^*(i_h, i'_{h-1}) + \sum_{j^* \text{ in the pref list of } i_h} w^*(i_h, j^*)$$

$$= 1 - \sum_{l=1}^{hc} w(i, j_l) + \sum_{l=1}^{(h-1)c} w(i, j_l) + \sum_{j^* \text{ in the pref list of } i_h} w(i, j)$$

$$= 1 - \sum_{l=(h-1)c+1}^{hc} w(i, j_l) + \sum_{l=(h-1)c+1}^{hc} w(i, j_l)$$

$$= 1$$

Finally, check condition 3. From above, we know that $w^*(i_h, i'_{h-1}) = w^*(i'_{h-1}, i'_{h-1})$, so the first element in each preference list in the new game (the first preference of $i_h$ is for $i'_{h-1}$) will always obey $w^*(i, a) = w^*(a, a)$. Also from above, $w^*(j^*, j^*) = w(j, j)$ and $w^*(i_h, j^*) = w(i, j)$. Therefore, if any lower preference disobeys $w^*(i, a) = w^*(a, a)$, then it must also be true that $w(i, a) \neq w(a, a)$. Since we assumed the $w$ values were an equilibrium in the original game, this must mean that there is no $b$ preferred less than $a$ with $w(i, b) > 0$, so for all $b^*$ preferred less than $a^*$, $w^*(i_h, b^*) = 0$.

**Equilibrium in the new preference game maps to an equilibrium in the original preference game.**

This map is simple. Given weights $w^*$ in the new preference game, create weights $w$ in the original preference game as follows.

- $w(i, j) = \max_h w^*(i_h, j^*)$

- $w(i, i) = w^*(i^*, i^*)$

The max in the first rule is a notational shortcut, since only one of the $i_h$ players will have any preference for $j^*$, and therefore at most one of the $i_h$ players will have $w^*(i_h, j^*) > 0$.

As before, we need to show the following to verify that this is an equilibrium in the original game.

1. $w(i, j) \leq w(j, j)$ for all $i, j$.

2. $w(i, i) + \sum_{j \neq i} w(i, j) = 1$ for all $i$.

3. If $w(i, j) > 0$, and if $i$ prefers $a$ over $j$, then $w(i, a) = w(a, a)$.

To show condition 1, consider players $i$ and $j$. Let $i_h$ = the player in the new game that has $j^*$ in its preference list. Now, $w(i,j) = w^*(i_h, j^*)$ and $w(j,j) = w^*(j^*, j^*)$. Since $w^*$ was feasible, we know that $w^*(i_j, j^*) \leq w^*(j^*, j^*)$, as desired.

Next, to show condition 2, consider player $i$. $w(i,i) + \sum_{x=1}^{k} w(i, j_x) = w^*(i^*, i^*) + \sum_{x=1}^{k} \max_h w^*(i_h, j_x^*)$. Let's compute $w^*(i^*, i^*)$ in the new preference game. Recall, the preference list for player $i_h$ is $i'_{h-1}$, $j^*_{(h-1)c+1}$, $j^*_{(h-1)c+2}$, ..., $j^*_{hc}$, $i_h$, and specifically, the preference list for $i^*$ $(= i_d)$ is $i'_{d-1}$, $j^*_{(d-1)c+1}$, $j^*_{(d-1)c+2}$, ..., $j^*_k$, $i_d$. Thus,

$$
\begin{aligned}
w^*(i^*, i^*) &= 1 - w^*(i_d, i'_{d-1}) - \sum_{x=(d-1)c+1}^{k} w^*(i_d, j_x^*) \\
&= 1 - w^*(i'_{d-1}, i'_{d-1}) - \sum_{x=(d-1)c+1}^{k} \max_h w^*(i_h, j_x^*) \\
&= 1 - [1 - w^*(i'_{d-1}, i_{d-1})] - \sum_{x=(d-1)c+1}^{k} \max_h w^*(i_h, j_x^*) \\
&= w^*(i'_{d-1}, i_{d-1}) - \sum_{x=(d-1)c+1}^{k} \max_h w^*(i_h, j_x^*) \\
&= w^*(i_{d-1}, i_{d-1}) - \sum_{x=(d-1)c+1}^{k} \max_h w^*(i_h, j_x^*) \\
&= 1 - [w^*(i_{d-1}, i'_{d-2}) + \sum_{x=(d-2)c+1}^{(d-1)c} w^*(i_{d-1}, j_x^*)] - \sum_{x=(d-1)c+1}^{k} \max_h w^*(i_h, j_x^*) \\
&= 1 - w^*(i'_{d-2}, i'_{d-2}) - \sum_{x=(d-2)c+1}^{(d-1)c} \max_h w^*(i_h, j_x^*) - \sum_{x=(d-1)c+1}^{k} \max_h w^*(i_h, j_x^*) \\
&= 1 - w^*(i'_{d-2}, i'_{d-2}) - \sum_{x=(d-2)c+1}^{dc} \max_h w^*(i_h, j_x^*) \\
&= \ldots \\
&= 1 - w^*(i'_1, i'_1) - \sum_{x=c+1}^{k} \max_h w^*(i_h, j_x^*) \\
&= 1 - [1 - w^*(i'_1, i_1)] - \sum_{x=c+1}^{k} \max_h w^*(i_h, j_x^*) \\
&= w^*(i_1, i_1) - \sum_{x=c+1}^{k} \max_h w^*(i_h, j_x^*) \\
&= 1 - \sum_{x=1}^{c} w^*(i_1, j_x^*) - \sum_{x=c+1}^{k} \max_h w^*(i_h, j_x^*)
\end{aligned}
$$

$$
\begin{aligned}
&= 1 - \sum_{x=1}^{c} \max_{h} w^*(i_h, j_x^*) - \sum_{x=c+1}^{k} \max_{h} w^*(i_h, j_x^*) \\
&= 1 - \sum_{x=1}^{k} \max_{h} w^*(i_h, j_x^*)
\end{aligned}
$$

Putting this back into our sum for player $i$, we get

$$
\begin{aligned}
w(i,i) + \sum_{x=1}^{k} w(i, j_x) &= w^*(i^*, i^*) + \sum_{x=1}^{k} \max_{h} w^*(i_h, j_x^*) \\
&= 1 - \sum_{x=1}^{k} \max_{h} w^*(i_h, j_x^*) + \sum_{x=1}^{k} \max_{h} w^*(i_h, j_x^*) = 1.
\end{aligned}
$$

So 2 holds.

Finally, we need to verify if $w(i, j) > 0$, and if $i$ prefers $a$ over $j$, then $w(i, a) = w(a, a)$. If $w(i, j) > 0$, then $\max_h w^*(i_h, j^*) > 0$. Let $h =$ the $h$ that satisfies $\max_h w^*(i_h, j^*)$. Now, if $i$ prefers $a$ over $j$, then either (Case 1) $i_h$ prefers $a^*$ over $j^*$ or (Case 2) there is some $b < h$ such that $i_b$ has preference for $a^*$. Start with Case 1. Since we know that $w^*$ is an equilibrium for the new preference game, it must be true that $w^*(i_h, a^*) = w^*(a^*, a^*)$, so $w(i, a) = w(a, a)$, as desired.

For Case 2, since $w^*(i_h, j^*) > 0$, we know that for all $b < h$, $w^*(i_b', i_b') < 1$ (otherwise, for all $c > b$, $i_c$ would put weight 1 on $i_{c-1}'$, leaving no weight left for itself, so $i_c'$ would also be 1. Therefore, $i_h$ would put weight 1 on $i_{h-1}$, leaving no weight for $j^*$.) Therefore, for the $b$ with preference for $a^*$, $w^*(i_b', i_b') < 1 \Rightarrow w^*(i_b', i_b) > 0 \Rightarrow w^*(i_b, i_b) > 0$. Therefore, for all $c^*$ in the preference list for $i_b$ (including $a^*$), $w^*(i_b, c^*) = w^*(c^*, c^*)$. So $w^*(i_b, a^*) = w^*(a^*, a^*)$, so $w(i, a) = w^*(i_b, a^*) = w^*(a^*, a^*) = w(a, a)$, as desired.

**Reducing to a preference game with constant in-degree (at most 2), with $O(m'n)$ players.**

Suppose we have a player $i$ who exists in the preference lists of $m'$ other players: $j_1, j_2, \ldots, j_{m'}$. We will add extra players $i_1', i_1, i_2', i_2, \ldots, i_{m'}', i_{m'}$. The preference lists for these new players will be: $i_1'$ has list $(i, i_1')$, for all $k > 1$: $i_k'$ has list $(i_{k-1}, i_k')$, for all $k$, $i_k$ has list $(i_k', i_k)$. If $i$ plays itself with weight $v$, then $i_k'$ will play itself with weight $1 - v$ and $i_k$ will play itself with weight $v$. Each of these new players has in-degree 1 and out-degree 1. Now, we can replace $i$ with $i_k$ in the preference list for $j_k$, so that $i$ now has in-degree 1 and each $i_k$ has in-degree 2. This does not affect the degree of any $j_k$. □

## 4 Personalized Equilibria

In this section, we introduce a new notion of equilibrium for matrix games, in which a player may individually match her strategies to her opponents strategies without obeying a product distribution. Since this equilibrium allows different players to simultaneously choose different matchings across the strategies, we call this a *personalized equilibrium*. In Section 4.2, we characterize the set of all personalized equilibria in a $k$-player game. In Section 4.3, we show that finding a personalized equilibrium is **PPAD**-complete.

Suppose we are given a $k$-player matrix game between players $1, \ldots, k$. Each player $i$ has strategy set $S_i$. We are also given a utility function for each $i$ specified by $u_i : E \to \mathbb{R}$, where $E = \prod_j S_j$. Now, given probability distributions $p_j(S_j)$ for each $j \neq i$, a best response for player $i$ (when using traditional Nash payoffs) is defined by the $p_i(S_i)$ that satisfies the following, where $w$ is a weight function over $e \in E$.

$$\max \sum_{e \in E} w(e) u_i(e)$$
$$w(e) = \prod_{s \in e \cap S_j} p_j(s) \quad \text{for all } e \in E$$
$$w(e) \geq 0 \quad \text{for all } e \in E$$

The correlator in a correlated equilibrium [3] relaxes the requirement that $w$ be a product distribution; however, $w$ does satisfy, among other conditions, the projection constraint $\sum_{e:s \in e} w(e) = p_j(s)$ for all $s \in S_j, 1 \leq j \leq k$. For a personalized equilibrium, we further relax this by allowing each player to define her own weight function, $w_i$, so that in the best response of player $i$, $p_i(s)$ (and $w_i(e)$) satisfy the following.

$$\max \sum_{e \in E} w_i(e) u_i(e)$$
$$\sum_{e:s \in e} w_i(e) = p_j(s) \quad s \in S_j, 1 \leq j \leq k$$
$$w_i(e) \geq 0 \quad e \in E$$

We can view a matrix game as a hypergraph with nodes $V = \cup_j S_j$ and edges $E = \prod_j S_j$. Then, if we interpret the $p_j(s)$ values as capacities on the nodes and the utility function for player $i$ as weights on the edges from the perspective of player $i$, a personalized equilibrium is simultaneously a maximum-weight fractional hypergraph matching for each player.

The description of the game above is exponential in the number of players since we require that every edge connects one strategy of each player. To allow for more succinct descriptions, we generalize the game as follows. For each player $i$, we introduce a hypergraph with nodes $V = \cup_j S_j$ and edges $E_i$. The set $E_i$ is required to satisfy two conditions (that are satisfied by $E$): (i) for each $e$ in $E_i$ and player $j$, $e$ contains at most one element of $S_j$; (ii) there do not exist distinct $e$ and $e'$ in $E_i$ such that $e \subset e'$. In the game, player $i$ places a weight $w_i(e)$ on each edge in $E_i$. A player must still place a total of weight 1 on all her edges, and all weights must be non-negative. Since the edges of $E_i$ may not connect all players, however, we relax the projection constraint to $\sum_{e:s \in e} w_i(e) \leq p_j(s)$. Thus, the collection of weights $w_i(e), e \in E_i$, and probability distributions $p_i(s), s \in S_i$, over all players $i$, form a personalized equilibrium if for each $i$, $w_i(e)$ and $p_i(s)$ maximize $\sum_{e \in E_i} w_i(e) u_i(e)$ subject to the following constraints.

$$\sum_{e:s \in e} w_i(e) \leq p_j(s) \quad \forall s \in S_j, \forall j \neq i \tag{1}$$
$$\sum_{e:s \in e} w_i(e) = p_i(s) \quad \forall s \in S_i$$
$$w_i(e) \geq 0 \quad \forall e \in E_i$$

> PERSONALIZED EQUILIBRIUM: Given players $1\ldots k$, strategy set $S_i$, edge set $E_i$, and utility function $u_i : E_i \to \mathbb{R}$ for each player $i$. Find a probability distribution $p_i : S_i \to \mathbb{R}$ and a weight assignment $w_i : E_i \to \mathbb{R}$ for each player $i$ that obeys the constraints of LP 1 and maximizes $\sum_{e \in E_i} w_i(e) u_i(e)$.

Just as mixed Nash equilibria exist for every matrix game, we show that every game thus defined has a personalized equilibrium.

**Theorem 4.1.** *For every multi-player matrix game, a personalized equilibrium always exists.*

*Proof.* Given the matrix game $G$, we construct the $k$-player game $\mathcal{G}$ in which the $i$th player's strategy space is the set of all probability distribution functions over $S_i$ and the payoff is given by the personalized payoff function defined above. We can view the strategy space as the set of probability distribution functions over $S_i$ instead of weight assignments to $E_i$ since a weight assignment uniquely defines a probability distribution function, and since the payoffs and responses of the other players only depend on the $p_i(s)$ values, not on the $w_i(e)$ values. Then a personalized equilibrium of $G$ is equivalent to a Nash equilibrium of $\mathcal{G}$. By [38, Proposition 20.3], a game has a pure Nash equilibrium if the strategy space of each player is a compact, non-empty, convex space, and the payoff function of each player is continuous on the strategy space of all players and quasi-concave in the strategy space of the player. The set of probability distributions over $S_i$ is clearly nonempty, convex, and compact. Furthermore, given probability distributions $p_i$ over $S_i$, $1 \leq i \leq k$, the payoff for any player $i$ is simply the solution to the following linear program with variables $w_i(e)$, over $e \in E_i$.

$$\max \sum_{e \in E_i} w_i(e) u_i(e)$$
$$\sum_{e \in E_i : s \in e} w_i(e) \leq p_j(s) \quad s \in S_j, 1 \leq j \leq k$$
$$\sum_{e \in E_i} w_i(e) = 1 \qquad w_i(e) \geq 0 \qquad e \in E$$

It is easy to see that the payoff function is both continuous in the probability distributions of all players, and quasi-concave in the strategy space of player $i$, thus completing the proof of the theorem. □

We define PERSONALIZED EQUILIBRIUM as the problem of finding a personalized equilibrium in a given matrix game. $k$-PERSONALIZED EQUILIBRIUM is the same problem in a game with $k$ players for constant $k$. Note that the traditional definition of a graphical game [26] may be used in this setting with smaller edges. In $d$-GRAPHICAL PERSONALIZED EQUILIBRIUM, each player $i$ has a neighborhood $N_i$ of at most $d$ other players, and all edges defined for player $i$ are in $\prod_{j \in N_i} S_j$. Finally, we define $\epsilon$-APPROXIMATE PERSONALIZED EQUILIBRIUM as the problem of finding a set of weight assignments ($w_i(e) \geq 0$ is the weight assigned by player $i$ to edge $e$) such that (a) for every player $i$, $1 - \epsilon \leq \sum_e w_i(e) \leq 1$, (b) for each player pair $i$ and $j$, and for each strategy $s$, $\left|\sum_{e:s \in e} w_i(e) - \sum_{e:s \in e} w_j(e)\right| \leq \epsilon$, and (c) for any best response weight assignment $w_i^*$ for any player $i$, $\sum_e w_i^*(e) u_i(e) - \sum_e w_i(e) u_i(e) \leq \epsilon$.

## 4.1 Characterizing personalized equilibria in two player games

We can simplify the definition of personalized equilibria when discussing two player games. Consider a matrix game $(R, C)$ between two players ROW and COLUMN, in which player ROW has strategies $r_1, r_2, \ldots, r_m$ and player COLUMN has strategies $c_1, c_2, \ldots, c_n$. $R \in \mathbb{R}^{m \times n}$ is the payoff matrix of ROW, and $C \in \mathbb{R}^{m \times n}$ is the payoff matrix of COLUMN.

Like a standard bimatrix game, if player ROW selects $r_i$ and player COLUMN selects $c_j$, the payoff to ROW is $R[i,j]$ and the payoff to COLUMN is $C[i,j]$. Suppose ROW selects a distribution $x$ among the strategies $\{r_1, r_2, \ldots, r_m\}$, and COLUMN selects a distribution $y$ among $\{c_1, c_2, \ldots, c_n\}$. Unlike payoffs defined for mixed strategies, in which the payoff to ROW is $\sum_{i,j} x[i]y[j]R[i,j]$ and the payoff to COLUMN is $\sum_{i,j} x[i]y[j]C[i,j]$, we define the payoffs using flows. The payoff to ROW is:

$$\text{Payoff (ROW)} = \max_{u_{i,j}} \sum_{i,j} u_{i,j} R[i,j] \tag{2}$$

$$\textbf{subject to } \sum_j u_{i,j} = x[i], \quad \forall i \quad \text{and} \quad \sum_i u_{i,j} = y[j], \quad \forall j;$$

$$\text{Payoff (COLUMN)} = \max_{v_{i,j}} \sum_{i,j} v_{i,j} C[i,j] \tag{3}$$

$$\textbf{subject to } \sum_j v_{i,j} = x[i], \quad \forall i \quad \text{and} \quad \sum_i v_{i,j} = y[j], \quad \forall j.$$

In other words, Payoff (ROW) is the cost of a 1-unit min-cost flow from source $r$ to destination $c$ in the directed graph $G_R = (V_R, E_R)$, with

$$V_R = \{r, c, r_1, r_2, \ldots, r_m, c_1, c_2, \ldots, c_n\}$$
$$E_R = \{(r \to r_i), \forall i\} \cup \{(r_i \to c_j), \forall i, j\} \cup \{(c_j \to c), \forall j\},$$

where the capacity of edge $(r \to r_i)$ is $x[i]$, the capacity of edge $(c_j \to c)$ is $y[j]$, and the capacity of all other edges is $+\infty$. The cost of edge $(r_i \to c_j)$ is $-R[i,j]$, and the cost of all other edges is 0. We note that for any distributions $x$ and $y$, a unit-flow from $r$ to $c$ always exists, so the above payoff function is well-defined.

Similarly, Payoff (COLUMN) is the cost of a 1-unit minimum-cost flow from source $c$ to destination $r$ in the directed graph $G_C = (V_C, E_C)$, with

$$V_C = \{r, c, r_1, r_2, \ldots, r_m, c_1, c_2, \ldots, c_n\}$$
$$E_C = \{(c \to c_j), \forall j\} \cup \{(c_j \to r_i), \forall i, j\} \cup \{(r_i \to r), \forall i\},$$

where the capacity of edge $(c \to c_j)$ is $y[j]$, the capacity of edge $(r_i \to r)$ is $x[i]$, and the capacity of all other edges is $+\infty$. The cost of edge $(c_j \to r_i)$ is $-C[i,j]$, and the cost of all other edges is 0.

It is not hard to show that the set of all two-player personalized equilibria is convex. In fact, we can give a stronger characterization, which will lead to a polynomial time algorithm.

**Theorem 4.2.** *A 2-player personalized equilibrium can always be found in polynomial time.*

*Proof.* Let graph $G$ = the union of $G_R$ and $G_C$. We will now consider a subgraph $G' = (V', E') \subset G$, such that $V' = V_R \cap V_C$, $(r_i \to c_j) \in E_R$ is in $E'$ if and only if $R[i,j] \geq R[i',j]$ for all $i'$, and $(c_j \to r_i) \in E_C$ is in $E'$ if and only if $C[i,j] \geq C[i,j']$ for all $j'$.

**Any directed cycle in $G'$ corresponds to a personalized equilibria.** Consider any cycle $\{r_{i1}, c_{j1}, r_{i2}, c_{j2}, \ldots, r_{il}, c_{il}\}$ in $G'$, each node played with weight $\frac{1}{l}$. Player ROW can match each of his strategies $r_{ik}$ with player COLUMN's strategy $c_{jk}$. Since this is a best response for player ROW, ROW cannot do better by changing to another strategy. Similarly, player ROW can match each of his strategies $c_{jk}$ with player ROW's strategy $r_{i(k+1)}$ for $k < l$, $c_{jl}$ can be matched with $r_{i1}$.

**Every personalized equilibria is a linear combination of cycles in $G'$.** Starting with any bipartite graph from $G'$ in which the in-degree equals the out-degree of each node (a characteristic of any personalized equilibria), we can remove any cycle (which is a personalized equilibria) and we are still left with a bipartite graph with the same characteristic. $\square$

## 4.2 Characterizing personalized equilibria in $k$-player games

We have shown that the set of all personalized equilibria for a two-player game is just the set of all linear combinations of cycles in an appropriately defined graph, which is easy to compute in polynomial time. However, for $k$ player games ($k > 3$), we will give a reduction from finding an equilibrium in a preference game to finding a personalized equilibrium in a $k$ player game (for $k > 3$), thereby showing that finding personalized equilibria is **PPAD**-hard. Nevertheless, we are able to give a concise characterization of the set of all personalized equilibria for arbitrary multi-player games.

**Theorem 4.3** (Personalized Equilibrium Characterization). *The following program represents the set of all exact personalized equilibria. The variables are $w_i(e)$, the weight placed by player $i$ on edge $e$, $\forall e \in E_i$.*

$$
\begin{align}
\sum_{e \in E_i : s \in e} w_i(e) &\leq \sum_{e \in E_j : s \in e} w_j(e) \quad s \in S_j, 1 \leq j, i \leq k \tag{4}\\
\sum_{e \in E_i} w_i(e) &= 1 \quad 1 \leq i \leq k\\
w_i(e) &\geq 0 \quad 1 \leq i \leq k, e \in E_i\\
\min_{e \in F} w_i(e) &= 0 \quad \text{for all players } i \text{ and subsets } F \subseteq E_i \text{ such that LP (5) is feasible.}
\end{align}
$$

*The following linear program is defined for each player $i$ and $F \subseteq E_i$ (referred to as an* improvement set*). The variables are $\delta(e)$ for each edge $e \in E_i$.*

$$
\begin{align}
\sum_{e \in E_i} \delta(e) u_i(e) &> 0 \tag{5}\\
\sum_{e \in E_i : s \in e} \delta(e) &= 0 \quad s \in S_j, 1 \leq j \leq k, j \neq i\\
\delta(e) &< 0 \quad (e \in F)\\
\delta(e) &\geq 0 \quad (e \notin F)
\end{align}
$$

Before formally proving this theorem, we will start with some intuition about why this characterizes all equilibria. here we provide some intuition. The first two constraints of program 4 specify a feasible weight assignment, and the first two constraints of LP 5 specify feasible "weight changes" that would increase the payoff for player $i$. How do we know that checking this for all subsets of edges is enough to find any possible improvement, and how does the last constraint of program 4 ensure that no improvement is possible? We can think of the $\delta$ values found in any solution to LP 5 as an "improvement direction." This is a vector that is orthogonal to the vector of all 1's and has a positive dot product with the utilities of $i$. In other words, if player $i$ were to move weight in this direction, her payoff would improve. Of course, there may be a continuum of such improvement directions. However, there are most an exponential number of negative supports, or "improvement sets". These are exactly the $F$ values for which LP 5 is feasible. Given an improvement set, the associated player can get a higher payoff by removing weight from all of those edges and adding them instead to edges with positive $\delta$ value. This improvement will be possible unless the player does not have weight on this entire improvement set; that is, unless $\min_{e \in F} w_i(e) = 0$.

*Proof.* **A solution to the program is an exact personalized equilibrium**. Assume we have a solution to Equation 4 that is not a personalized equilibrium. The first two constraints ensure that our solution is a feasible weight assignment for the game. Therefore, there must be some player $i$ who is not playing a best response. Take some better response, in which player $i$ plays weights $w_i^*(e)$, and let $\delta(e) = w_i^*(e) - w_i(e)$.

Let $F$ be the subset of $E_i$ such that $\delta(e) < 0$ (that is, player $i$ puts more weight on each edge in $F$ in the original response than in the best response). Since $w_i^*$ has a strictly higher total utility for player $i$ than $w_i$, we know that $\sum_{e \in E_i} w_i^*(e) u_i(e) > \sum_{e \in E_i} w_i(e) u_i(e)$, which implies that $\sum_{e \in E_i} \delta(e) u_i(e) > 0$. Since both $w_i$ and $w_i^*$ were feasible weights, it must be true for any strategy $s$ that $\sum_{e:s \in e} w_i(e) = \sum_{e:s \in e} w_i^*(e) \Rightarrow \sum_{e:s \in e} \delta(e) = 0$.

By our definition of $F$, $\delta(e) < 0$ for all $e \in F$ and $\delta(e) \geq 0$ for all $e \notin F$. Therefore, $F$ and $i$ obey all the constraints of linear program 5, so since $w_i(e)$ obeyed program 4, we know that $\min_{e \in F} w_i(e) = 0$. Thus, there exists some edge $f \in F$ such that $w_i(f) = 0$. But then $0 > \delta(f) = w_i^*(f) - w_i(f) = w_i^*(f)$, contradicting the fact that $w_i^*$ was a feasible best response for player $i$.

**Any personalized equilibrium is a solution to the program**. Assume we have a personalized equilibrium that does not satisfy some constraint of the program. Let $w_i(e)$ be the weight placed by player $i$ on edge $e$ in this equilibrium. The first three constraints are the definition of a feasible weight assignment. Therefore, assume this equilibrium does not satisfy the min constraint for some player $i$ and some subset $F \subseteq E_i$ for which LP 5 is feasible.

Consider a solution $\delta$ for LP 5 for this $i$ and $F$. Let $M = \frac{\min_{e \in F} |w_i(e)|}{\max_{e \in F} |\delta(e)|}$, and let $\delta'(e) = \delta(e) \cdot M$. $M$ is a well-defined positive number since (1) the min constraint was not satisfied, and (2) LP 5 specifies that $\delta(e) < 0$ for all $e \in F$. We know that $F$ is non-empty because the first constraint implies there is some $\delta > 0$, and combining this with the second constraint implies there is also some $\delta < 0$. Furthermore, for all $e$ with $\delta(e) < 0$ (i.e., for all $e \in F$), $|\delta'(e)| \leq w_i(e)$. Now, consider the alternate assignment for player $i$ specified by $w_i^*(e) = w_i(e) + \delta'(e)$. By the second constraint of LP 5 and the fact that $|\delta'(e)| \leq w_i(e)$ for all $e$ with $\delta(e) < 0$, this is still a valid weight assignment. By the first constraint of LP 5, this gives a strictly higher total utility for player $i$. Therefore, weights $w_i(e)$ did not give a best response for player $i$, so we did not have a personalized equilibrium, contradicting our assumption and completing the proof. □

**Corollary 4.4.** *For any matrix game with all rational payoffs, there exists a personalized equilibrium in which the probability assigned by each player to each strategy is a rational number.*

*Proof.* In Theorem 4.3, we showed that any personalized equilibrium is a solution to a linear program plus additional min constraints, in which all coefficients are rational. By Theorem 4.1, this program has at least one solution. Now, we can rewrite this as a union of many linear programs as follows. Let $F_1, \ldots, F_\alpha$ be the set of all improvement sets. We can write $\prod_{i=1}^{\alpha} |F_i|$ linear programs, each consisting of the first three constraints from program 4 as well as the $\alpha$ constraints $[e_1 = 0$ for some $e_1 \in F_1]$, $[e_2 = 0$ for some $e_2 \in F_2]$, ..., $[e_\alpha = 0$ for some $e_\alpha$ in $F_\alpha]$. We can create one LP for each such combination of one edge from each improvement set, or $\prod_{i=1}^{\alpha} |F_i|$ LPs. Since the union of these linear programs is exactly the same as the program in Theorem 4.3, and since (by Theorem 4.1) the program in Theorem 4.3 has at least one solution, we know that at least one of these linear programs has a solution. Any feasible LP with rational coefficients will have a rational solution. Therefore, there will be a personalized equilibrium with all rational weights. □

## 4.3 Finding personalized equilibria is PPAD-complete

This section contains four reductions. First, we reduce DEGREE $d$ PREFERENCE GAME to $d$-GRAPHICAL PERSONALIZED EQUILIBRIUM. We next reduce 3-GRAPHICAL PERSONALIZED EQUILIBRIUM to 4-PERSONALIZED EQUILIBRIUM. It can be easily verified that the same reductions can be used to show $\epsilon$-APPROXIMATE PREFERENCE GAME $\leq_P$ $\epsilon$-APPROXIMATE PERSONALIZED EQUILIBRIUM. These reductions together show that finding an $\epsilon$-approximate personalized equilibrium in both graphical games and 4-player games is **PPAD**-hard.

**Theorem 4.5.** PREFERENCE GAME $\leq_P$ PERSONALIZED EQUILIBRIUM

*Proof.* Given a preference game over player set $[n]$, with the preference lists specified as a set of values $Q_{ij}$ for all $i, j \in [n]$: $Q_{ij}$ = the number of players $k$ such that $j \succeq_i k \succeq_i i$.

Define a game as follows, in which we will find a personalized equilibrium.

- The set of players = $\{p_1, \ldots, p_n\}$
- $S_i$ (the set of strategies for player $p_i$) = $\{s_{ij} : Q_{ij} > 0\}$
- $H_i$ = the set of hyperedges for player $p_i$ = $\{\{s_{ij}, s_{jj}\} \forall s_{ij} \in S_i, j \neq i\} \cup \{s_{ii}\}$
- $u_i(\{s_{ij}, s_{jj}\})$ (the payoff to player $i$ for this hyperedge) = $Q_{ij}$
- $u_i(\{s_{ii}\}) = Q_{ii} \geq 1$

Notice that the degree of the game is preserved, and the number of edges defined is at most $n$ times the degree.

**A personalized equilibrium maps to an equilibrium in the preference game.**
The map will be as follows: Suppose we are given weights $x_{ij}$ for each player $i$ and edge $\{s_{ij}, s_{jj}\}$, and $x_{ii}$ for player $i$ and hyperedge $\{s_{ii}\}$. These weights form a personalized equilibrium. We will set weights $w_{ij} = x_{ij}$ in the preference game.

To show this is an equilibrium in the preference game, we must show the following.

- For all $i$, $\sum_j w_{ij} = 1$.

    $\sum_j w_{ij} = \sum_j x_{ij} = 1$, since this is a valid solution to the personalized game.

- For all $i, j$, $w_{ij} \leq w_{jj}$.

    $w_{ij} = x_{ij} \leq x_{jj}$ (by the projection constraint for personalized equilibria), $x_{jj} = w_{jj}$.

- $w$ is a lexicographically maximal weight assignment.

  Suppose this is not true. Then, there exists another weight assignment $w'$ that is lexicographically larger than $w$. Let $w'$ be the lexicographically maximal such assignment. Thus, there exist $i, j$ such that $\sum_{k:Q_{ik} \geq Q_{ij}} w'_{ij} > \sum_{k:Q_{ik} \geq Q_{ij}} w_{ij}$. For this $i$, take a $j$ with the largest $Q_{ij}$ that meets this condition. By our definition of $j$ and the fact that $w'$ is lexicographically maximal, we know that for all $j'$ with $Q_{ij'} > Q_{ij}$, $\sum_{k:Q_{ik} \geq Q_{ij'}} w'_{ij} = \sum_{k:Q_{ik} \geq Q_{ij'}} w_{ij}$. Let $\delta = \sum_{k:Q_{ik} \geq Q_{ij}} w'_{ij} - \sum_{k:Q_{ik} \geq Q_{ij}} w_{ij} > 0$. Now consider the payoff to player $i$ in the personalized game.

$$
\begin{aligned}
\text{Payoff to } i &= \sum_l x_{il} Q_{il} \\
&= \sum_{l:Q_{il}>Q_{ij}} x_{il} Q_{il} + \sum_{l:Q_{il}=Q_{ij}} x_{il} Q_{il} + \sum_{l:Q_{il}<Q_{ij}} x_{il} Q_{il} \\
&= \sum_{l:Q_{il}>Q_{ij}} w_{il} Q_{il} + Q_{ij} \sum_{l:Q_{il}=Q_{ij}} w_{il} + \sum_{l:Q_{il}<Q_{ij}} w_{il} Q_{il} \\
&= \sum_{l:Q_{il}>Q_{ij}} w'_{il} Q_{il} + Q_{ij} \sum_{l:Q_{il}=Q_{ij}} w_{il} + \sum_{l:Q_{il}<Q_{ij}} w_{il} Q_{il} \\
&= \sum_{l:Q_{il}>Q_{ij}} w'_{il} Q_{il} + Q_{ij}((\sum_{l:Q_{il}=Q_{ij}} w'_{il}) - \delta) + \sum_{l:Q_{il}<Q_{ij}} w_{il} Q_{il} \\
&\leq \sum_{l:Q_{il}>Q_{ij}} w'_{il} Q_{il} + Q_{ij} \sum_{l:Q_{il}=Q_{ij}} w'_{il} - \delta Q_{ij} + \sum_{l:Q_{il}<Q_{ij}} w'_{il} Q_{il} + \delta(Q_{ij}-1) \\
&< \sum_{l:Q_{il}>Q_{ij}} w'_{il} Q_{il} + Q_{ij} \sum_{l:Q_{il}=Q_{ij}} w'_{il} - \delta(Q_{ij}-1) + \sum_{l:Q_{il}<Q_{ij}} w'_{il} Q_{il} + \delta(Q_{ij}-1) \\
&= \sum_{l:Q_{il}>Q_{ij}} w'_{il} Q_{il} + Q_{ij} \sum_{l:Q_{il}=Q_{ij}} w'_{il} + \sum_{l:Q_{il}<Q_{ij}} w'_{il} Q_{il} \\
&= \sum_l w'_{il} Q_{il}
\end{aligned}
$$

So player $i$ would do strictly better by playing $x = w'$, leading to a contradiction.

**An equilibrium in the preference game maps to a personalized equilibrium.**

Suppose we are given weights $w_{ij}$ forming an equilibrium in the preference game. We will set weights in the personalized game as follows. $x_{ih} = w_{ij}$ for player $i$ and edge $h = \{s_{ij}, s_{jj}\}$. $x_{ih} = w_{ii}$ for player $i$ and edge $h = \{s_{ii}\}$.

To show this is a personalized equilibrium, we must show the following.

- For all $i$, $\sum_h x_{ih} = 1$.

  $\sum_h x_{ih} = \sum_j w_{ij} = 1$, since this is a valid weight assignment in the preference game.

- For all $i, s_{jk} \in S_j$, $\sum_{h:s_{jk} \in h} x_{ih} \leq \sum_{h:s_{jk} \in h} x_{jh}$.

  If $j \neq k$, $\sum_{h:s_{jk} \in h} x_{ih} = 0 \leq \sum_{h:s_{jk} \in h} x_{ij}$. If $j = k$, $\sum_{h:s_{jj} \in h} x_{ih} = x_{ih'}$ where $h' = \{s_{ij}, s_{jj}\}$ $= w_{ij} \leq w_{jj} = x_{jh''}$ where $h'' = \{s_{jj}\} = \sum_{h:s_{jj} \in h} x_{jh}$.

- $x$ is a best response in the personalized game for all players $i$.

    Consider any other weight function $x'$ for the personalized game. Since there is a one-to-one mapping from defined edges to $i, j$ pairs in the preference game (including $i = j$), we can define a new weight function $w'$ in the preference game using the same rules as defined in the first half of this proof ($w'_{ij} = x'_{ij}$). We know that $w$ is lexicographically maximal for the preference game. Using the same reasoning as above, we get:

$$\begin{aligned}
\text{Payoff to } i \text{ playing } x' &= \sum_l x'_{il} Q_{il} \\
&= \sum_{l:Q_{il}>Q_{ij}} x'_{il} Q_{il} + \sum_{l:Q_{il}=Q_{ij}} x'_{il} Q_{il} + \sum_{l:Q_{il}<Q_{ij}} x'_{il} Q_{il} \\
&= \sum_{l:Q_{il}>Q_{ij}} w'_{il} Q_{il} + Q_{ij} \sum_{l:Q_{il}=Q_{ij}} w'_{il} + \sum_{l:Q_{il}<Q_{ij}} w'_{il} Q_{il} \\
&< \sum_{l:Q_{il}>Q_{ij}} w_{il} Q_{il} + Q_{ij} \sum_{l:Q_{il}=Q_{ij}} w_{il} + \sum_{l:Q_{il}<Q_{ij}} w_{il} Q_{il} \\
&= \sum_l w_{il} Q_{il} \\
&= \text{Payoff to } i \text{ playing } x
\end{aligned}$$

□

**Corollary 4.6.** *It is* **PPAD**-*hard to find a personalized equilibrium, even in a graphical game with degree 3.*

*Proof.* The reduction in the proof to Theorem 4.5 preserves the number of players. For player $i$, it creates one hyperedge of size at most 2 for each player in out($i$) in the preference game. Therefore, the degree in the game is preserved. □

**Theorem 4.7.** 3-GRAPHICAL PERSONALIZED EQUILIBRIUM $\leq_P$ 4-PERSONALIZED EQUILIBRIUM

*Proof.* Suppose we are given a graphical game with degree 3 for which we want to find a personalized equilibrium. We first convert this graph so that it still has degree 3 but also obeys the following property: For each node $u$ with at least two outgoing edges, one to node $v_1$ and the other to node $v_2$, there exists either an edge from $v_1$ to $v_2$ or an edge from $v_2$ to $v_1$.

Given a graph with maximum degree 3, we make the following modifications to satisfy the above property. Suppose we have a node $u$ with an edge to $v_1$ and an edge to $v_2$, and suppose $v_1$ already has degree 3. To fix this, we will create an extra node $v'_1$ with the same strategies as $v_1$ which will play exactly the same weights as $v_1$ in an equilibrium. We can do this by setting the payoff to $v'_1$ for agreeing with $v_1$ to 1, and the payoffs for disagreeing with $v_1$ to 0. $v'_1$ has out-degree 1, since it only depends on $v_1$. We have added one to the in-degree of $v_1$. However, now we can replace edge $(u, v_1)$ with edge $(u, v'_1)$, so $v_1$ now has degree 3, as originally, and $v'_1$ has degree 2, so we can add an edge between $v'_1$ and $v_2$ without exceeding the degree requirement on $v_1$. Repeating the above transformations with other vertices that violate the desired property will lead to a degree-3 graphical game that satisfies the property.

Now we have a degree 3 graph with the desired property. Create a 3-coloring and create one player per color. Each player takes each of the strategies for each node in that color. For ease of notation, assume that each of the original nodes had only 2 strategies. This can be easily adjusted for more strategies. Add dummy strategies as necessary so that each of the 3 players has the same number of strategies. Also add a fourth player with half the number of strategies as any other player.

This gives us 4 players. Let the strategies for player 1 be $\{a_{10}, a_{11}, a_{20}, a_{21}, \ldots, a_{k0}, a_{k1}\}$. The strategies for player 2 are $\{b_{10}, b_{11}, \ldots, b_{k0}, b_{k1}\}$. The strategies for player 3 are $\{c_{10}, c_{11}, \ldots, c_{k0}, c_{k1}\}$. The strategies for player 4 are $\{d_1, d_2, \ldots, d_k\}$.

Next we will assign payoffs for each hyperedge. Start by giving each hyperedge the same payoff as in the graphical game (we can do this because no two nodes influencing the same strategy are strategies of the same player). Notice that these payoffs will not depend at all on player 4. All of player 4's payoffs start at 0. Let $p_i(w, x, y, z) = $ the payoff to player $i$ if player 1 plays $w$, 2 plays $x$, player 3 plays $y$, player 4 plays $z$. Now we want to add to these payoffs in order to ensure that each player plays each strategy pair equally.

Let $M$ be strictly greater than the largest payoff so far. Now, change the following payoffs:

$p_1(a_{si}, x, y, d_s) += M$ (player 1 is playing either strategy from the node numbered $s$, player 4 is playing his $s^{th}$ strategy).
$p_2(w, b_{si}, y, d_s) += M$ (player 2 is playing either strategy from the node numbered $s$, player 4 is playing his $s^{th}$ strategy).
$p_3(w, x, c_{si}, d_s) += M$ (player 3 is playing either strategy from the node numbered $s$, player 4 is playing his $s^{th}$ strategy).
$p_4(a_{si}, x, y, d_{(s+1)}) += M$ (player 1 is playing either strategy from the node numbered $s$, player 4 is playing strategy $s \bmod k + 1$).

If $f_i(x) = $ the amount player $i$ plays strategy $x$ then in any equilibrium we must have (for all $s$)

$$\begin{aligned}
f_1(a_{s0}) + f_1(a_{s1}) &= f_4(d_s) \\
f_2(b_{s0}) + f_2(b_{s1}) &= f_4(d_s) \\
f_3(c_{s0}) + f_3(c_{s1}) &= f_4(d_s) \\
f_4(d_s) &= f_1(a_{(s-1)0}) + f_1(a_{(s-1)1}) \text{ for } 1 < s \leq k \\
f_4(d_1) &= f_1(a_{k0}) + f_1(a_{k1})
\end{aligned}$$

These equations imply that:

$$\begin{aligned}
f_1(a_{s0}) + f_1(a_{s1}) &= f_1(a_{(s-1)0}) + f_1(a_{(s-1)1}) \text{ for } s > 0 \\
f_1(a_{00}) + f_1(a_{01}) &= f_1(a_{k0}) + f_1(a_{k1}) \text{ for } s > 0 \\
f_2(b_{s0}) + f_2(b_{s1}) &= f_1(a_{s0}) + f_1(a_{s1}) \\
f_3(c_{s0}) + f_3(c_{s1}) &= f_1(a_{s0}) + f_1(a_{s1})
\end{aligned}$$

In other words, given a personalized equilibrium in this game, we can simply multiply by the number of pairs (nodes) per player to get a personalized equilibrium in the graphical game. □

The remaining two reductions will be used to show **PPAD** membership of Personalized Equilibrium. We show how to reduce Personalized Equilibrium to $\epsilon$-Approximate Personalized Equilibrium, as long as $\epsilon$ is sufficiently small. Finally, we reduce $\epsilon$-Approximate Personalized Equilibrium to End of the Line, thereby completing the proof that finding personalized equilibria (as well as $\epsilon$-approximate personalized equilibria) is **PPAD**-complete. We start with an LP compactness claim that will be useful to show Personalized Equilibrium $\leq_P$ $\epsilon$-Approximate Personalized Equilibrium.

**Lemma 4.8** (LP Compactness). *If an LP with $n$ variables and rational coefficients, each represented by at most $\beta$ bits, is such that there is a point obeying each constraint to within $\varepsilon = \frac{1}{2^{3n\beta}}$, then the LP is feasible.*

*Proof.*

**Lemma 4.9.** *If $t$, $b$, $t_i$, $b_i$, $y_i$, $z_i$, (for $1 \leq i \leq n$), are $\beta$-bit integers, then either $\sum_{i=1}^n \frac{t_i y_i}{b_i z_i} \geq \frac{t}{b}$ or $\sum_{i=1}^n \frac{t_i y_i}{b_i z_i} < \frac{t}{b} - \frac{1}{2^{3n\beta}}$.*

*Proof.* Suppose we have $\sum_{i=1}^n \frac{t_i y_i}{b_i z_i} < \frac{t}{b}$. Then the difference $\frac{t}{b} - \sum_{i=1}^n \frac{t_i y_i}{b_i z_i}$ is at least $1/(b \cdot \prod_i b_i \prod_i z_i)$, which is at least $1/2^{\beta+2n\beta} < 2^{-3n\beta}$ since each integer in the product is at most $2^\beta$. □

From Lemma 4.9, we get the following. If $x$ satisfies $\sum_{i=1}^n a_i x_i \geq b - \frac{1}{2^{3n\beta}}$, where each $a_i$ and $b$ are rational numbers whose numerators and denominators are representable as $\beta$-bit integers, then $x$ satisfies $\sum_{i=1}^n a_i x_i \geq b$. This immediately implies the Lemma. □

**Corollary 4.10.** *Given a linear program with $\leq n$ variables and coefficients of the form $\frac{a}{b}$ for integers $a$ and $b$, each represented by at most $\beta$ bits, each coordinate of a vertex must be representable by $\frac{c}{d}$ for integers $c$ and $d$, each represented by less than $3n\beta$ bits.*

We use the next lemma to show that an approximate personalized equilibrium *almost* satisfies the constraints of Theorem 4.3. As long as $\epsilon$ is small enough, this will imply by Lemma 4.8 that an $\epsilon$-approximate personalized equilibrium will point us to a feasible LP that finds an exact personalized equilibrium.

**Lemma 4.11.** *An $\epsilon_1$-approximate personalized equilibrium (with $\epsilon_1 = \frac{1}{2^{3|E|(\beta+\gamma)}}$) will obey every constraint in program 4 to within $\epsilon_2 = \frac{1}{d}$ for $\gamma$ bit integer $d$, if each utility value is representable as $\frac{a}{b}$ for integers $a$ and $b$ of at most $\beta$ bits each.*

*Proof.* Assume for the sake of contradiction that we have an $\epsilon_1$-approximate personalized equilibrium that does not satisfy some constraint of the program to within $\epsilon_2$. Let $w_i(e)$ = the weight placed by player $i$ on hyperedge $e$ in this approximate equilibrium. The first three constraints of program 4 must be satisfied to within $\epsilon_1$, since they are the definition of a feasible weight assignment. Therefore, assume this equilibrium does not satisfy the min constraint to within $\epsilon_2$ for some player $i$ and some subset $F \subseteq E_i$ for which LP 5 is feasible.

Consider a solution $\delta^*$ for LP 5 for this $i$ and $F$. Let $M = \frac{\epsilon_2}{\max_{e \in F} |\delta^*(e)|}$, and let $\delta'(e) = \delta^*(e) \cdot M$. $M$ is well-defined since $\delta^*(e) < 0$ for all $e \in F$. Furthermore, for all $e$ with $\delta(e) < 0$ (i.e., for all $e \in F$), $|\delta'(e)| \leq \epsilon_2 \leq w_i(e)$.

Now consider the following slightly adjusted linear program.

$$\text{maximize} \sum_{e \in E_i} \delta(e) u_i(e)$$

$$\sum_{e:s\in e} \delta(e) = 0 \quad s \in S_j, 1 \leq j \leq k, j \neq i \qquad (6)$$

$$\delta(e) \geq -\epsilon_2 \quad (e \in F)$$

$$\delta(e) \geq 0 \quad (e \notin F)$$

By our choice of $\delta'$ and the analysis above, $\delta'$ obeys each constraint of the new linear program 6, and $\delta'$ gives a maximization value $> 0$. LP 6 has $|E_i| \leq |E|$ variables, each coefficient $u_i(e)$ can be represented as $\frac{a}{b}$ for $\beta$-bit integers $a$ and $b$, and coefficient $\epsilon_2 = \frac{1}{d}$ for $\gamma$ bit integer $d$. The maximization point of an LP will be at a vertex, so by Corollary 4.10, each dimension of the maximization point of LP 6 will be representable by $\frac{c}{b}$ for integers $c$ and $d$, each of less than $3|E|(\beta + \gamma)$ bits.

Let $\delta =$ the solution to LP 6, and consider the alternate assignment for player $i$ specified by $w_i^*(e) = w_i(e) + \delta(e)$. By the first constraint of LP 6, this still does not overfill or underfill any strategy by more than $\epsilon_1$, and player $i$ still places total weight between $1 - \epsilon_1$ and $1 + \epsilon_1$. By the second and third constraints, $\delta(e) < 0$ if and only if $e \in F$, in which case $|\delta(e)| \leq \epsilon_2$ (when $w_i(e) \geq \epsilon_2$), so $w_i^*(e) \geq 0$. In other words, we still have a valid weight assignment for an $\epsilon_1$-approximate personalized equilibrium. Since we know LP 6 has a solution $> 0$, and (from above) each coordinate of the solution is representable as $\frac{c}{d}$ for integers $c$ and $d$, each of at most $3|E|(\beta+\gamma)$, this gives a total utility for player $i$ that is more than $\frac{1}{2^{3|E|(\beta+\gamma)}} = \epsilon_1$ larger than the original total utility. Therefore, the original solution was not a valid $\epsilon_1$-approximate personalized equilibrium, contradicting our assumption and completing the proof. □

**Theorem 4.12.** *There is a polynomial time reduction from finding an exact personalized equilibrium in a game with $|E|$ edges to finding an $\epsilon$-approximate personalized equilibrium for any $\epsilon \leq \frac{1}{|E|^{3|E|}2^{9|E|^2\beta + 3|E|\beta}}$, assuming all utilities can be represented as $\frac{a}{b}$ for integers $a$ and $b$, each of at most $\beta$ bits.*

*Proof.* The reduction consists of two steps. In the first step, we find an $\epsilon$-approximate personalized equilibrium for the given game in which we want to find an exact personalized equilibrium. In the second step, we solve the following linear program for $w_i(e)$ and $p_i(s)$, $\forall i \in 1 \ldots k, s \in S_i, e \in E_i$. Let $E_i' \subset E_i =$ the set of all edges $e$ such that player $i$ assigned $\leq 1/(|E| \cdot 2^{3|E|\beta})$ weight to edge $e$ in the approximate equilibrium.

$$\sum_{e \in E_i: s \in e} w_i(e) \leq p_j(s) \quad 1 \leq i, j \leq k, s \in S_j \qquad (7)$$

$$\sum_{s \in S_i} p_i(s) = 1 \quad 1 \leq i \leq k$$

$$w_i(e) = 0 \quad \forall e \in E_i', 1 \leq i \leq k$$

$$w_i(e) \geq 0 \quad 1 \leq i \leq k, e \in E_i \setminus E_i'$$

$\epsilon \leq \frac{1}{E^{3|E|} * 2^{9|E|^2\beta + 3|E|\beta}} = \frac{1}{2^{3|E|\log_2|E| + 9|E|^2\beta + 3|E|\beta}} = \frac{1}{2^{3|E|(\beta + 3|E|\beta + \log_2|E|)}}$. Since $\epsilon \leq \frac{1}{2^{3|E|(\beta + 3|E|\beta + \log_2|E|)}}$, by Lemma 4.11, the approximate equilibrium found in the first step will satisfy all constraints of program 4 to within $\frac{1}{2^{3|E|\beta + \log_2|E|}} = 1/(|E| * 2^{3|E|\beta})$. Therefore, a solution found in the second step will exactly satisfy all "min" constraints in program 4. Clearly, a solution found in the second step will exactly satisfy all other constraints in program 4, since the other constraints are identical.

Each value $w_i(e)$ is being decreased by at most $1/(|E|2^{3|E|\beta})$ (but not to less than 0) from the approximate equilibrium to the solution to the second step, so the last two constraints in linear program 7 are satisfied exactly using the values from the approximate equilibrium, while the first two constraints in linear program 7 are satisfied to within $1/(2^{3|E|\beta})$ by the values from the approximate equilibrium. Therefore, the approximate equilibrium satisfies each constraint of linear program 7 to within $\epsilon = 1/(2^{3|E|\beta})$. By Lemma 4.8, linear program 7 is feasible, so by Lemma 4.3, we can find an exact personalized equilibrium. □

Theorem 4.12 and Theorem 4.13 put Personalized Equilibrium in **PPAD**.

**Theorem 4.13.** $\epsilon$-Approximate Personalized Equilibrium $\leq_P$ End of the Line

*Proof.* We used fixed point theorems to prove the existence of a personalized equilibrium, and relaxing the problem to finding $\epsilon$-approximate equilibria automatically moves us from a continuous to a discrete world. Here, we show that finding an $\epsilon$-approximate equilibrium is in **PPAD**. This is not surprising given that several discrete fixed point problems have been shown to be in the class **PPAD**. Our proof uses the machinery already established for proving that finding approximate Nash equilibrium in $r$-player games is in **PPAD** [13]. Their proof [13][Section 3.2] will apply to personalized equilibria as well, as long as we can define a polynomial-time computable function $f(x)$ for $x \in \mathbb{R}^{n \cdot k}$ for a game with $k$ players, $n$ strategies per player, which satisfies the conditions we enumerate below. We need to introduce some notation first. For $p \in [k]$ and $i \in [n]$, we denote by $x_p^i$ the $((p-1)n+i)$-th coordinate of $x$ (the amount that player $p$ plays strategy $i$). For player $p$, let $D_p$ denote the set $\{(p-1)n+j : j \in [n]\}$; that is, the dimensions corresponding to the strategies of $p$. Then $x_p$ is the projection of $x$ on $D_p$ and $x_{-p}$ is the projection of $x$ on $[nk] - D_p$. The function $f(x)$ must satisfy the following requirements for the proof to translate:

1. $\forall p, \sum_i f(x_p^i) = 1$

2. If $\|x - x'\|_\infty < \delta$, then $\|f(x) - f(x')\|_\infty < U_{\max} 2^{\text{poly}(n,m,k)} \delta$, where $m$ is the number of edges and $U_{\max}$ is the maximum payoff entry in the given instance.

3. If $\|f(x) - x\|_\infty < \epsilon_1$, then $x$ is an $\epsilon$-approximate personalized equilibrium. Here, we will use any $\epsilon_1 \leq \frac{\epsilon}{nU_{\max}}$. In the proof from [13], $\epsilon_1$ only affects the number of nodes in the End of the Line graph.

We define $f(x)$ as follows: we set $f(x)_p$ to be the lexicographically least best response to $x_{-p}$. We now show that $f$ satisfies the three conditions listed above. The first condition is immediate from the definition of $f(x)$. For the second condition, fix $x$, $x'$, and a player $p$. Then $f(x)_p$ is obtained by solving a best response linear program for player $p$ given the strategy distribution $x_{-p}$ of the other players. The LP, which we denote by $\mathcal{O}$ for this proof, is over the variables $f(x)_p^i$, for strategy $i \in S_p$, and $w_p(e)$ for every edge $e$, and maximizes a linear utility $u(f(x)_p, w_p)$ subject to linear constraints $B \cdot (f(x)_p, w_p)^T \geq c$. We note that every element of $B$ is either 0 or 1 and every element of $c$ is either 0, 1, or a coordinate of $x_{-p}$. Similarly, $f(x')_p$ is an optimal solution to an LP $\mathcal{O}'$, which maximizes $u(f(x')_p^i, w_p')$ subject to $B \cdot (f(x)_p, w_p')^T \geq c'$, where $c'$ is derived from $x'_{-p}$ in the same way as $c$ is derived from $x_{-p}$.

Let $U$ and $U'$ denote the optimal values of $\mathcal{O}$ and $\mathcal{O}'$. We first argue that if $\|x - x'\|_\infty \leq \delta$, then $|U - U'| \leq mU_{\max}(nk)!\delta$, where $m$ is the number of edges and $U_{\max}$ is the maximum payoff entry in the given game. We note that $x$ satisfies the constraints of $\mathcal{O}'$ to within $\delta$. We also know that $\mathcal{O}'$

is feasible. The number of variables and constraints in both $\mathcal{O}$ and $\mathcal{O}'$ are $n+m$ and $nk+m+1$. Therefore, by Lemma 4.14, there exists a point $y$ that satisfies the constraints of $\mathcal{O}'_p$ such that $\|y-x\|_\infty \leq (nk)!\delta$; here we use the fact that every entry in the constraint matrix and vector of $\mathcal{O}$ and $\mathcal{O}'$ is at most 1. Thus, the utility achieved by $y$ is at least $u(f(x)_p, w_p) - mU_{\max}(nk)!\delta$, yielding $U' \geq U - mU_{\max}(nk)!\delta$. Similarly, we have $U \geq U' - mU_{\max}(nk)!\delta$. This gives the desired bound $|U - U'| \leq mU_{\max}(nk)!\delta$.

By definition, we have that $(f(x)_p, w_p)$ is the lexicographically least element of the feasibility LP consisting of the constraints of $\mathcal{O}_p$ together with the constraint $u(f(x)_p, w_p) \geq U$. Let us call this LP $\mathcal{P}$. Similarly, $(f(x')_p, w'_p)$ is the lexicographically least element of the feasibility LP $\mathcal{P}'$ consisting of the constraints of $\mathcal{O}'$ together with the constraint $u(f(x')_p, w'_p) \geq U'$. We note that $P$ and $P'$ have the same set of variables and the same constraint matrix; that is, $P$ and $P'$ can be written down as $Ax \geq b$ and $Ax \geq b'$ respectively. Since $\|x - x'\|_\infty \leq \delta$ and $|U - U'| \leq mU_{\max}(nk)!\delta$, we have $\|b - b'\|_\infty \leq mU_{\max}(nk)!\delta$. We now apply Corollary 4.15 to obtain that $\|f(x)_p - f(x')_p\|_\infty$ is at most $U_{\max} 2^{\text{poly}(n,m,k)} \delta$.

For the third condition, recall our definition of an $\epsilon$-approximate personalized equilibrium. We require: (3a) for every player $p$, $1 - \epsilon \leq \sum_e w_p(e) \leq 1$, (3b) for each player pair $p$ and $q$, and for each strategy $s$, $|\sum_{e:s \in e} w_p(e) - \sum_{e:s \in e} w_q(e)| \leq \epsilon$, and (3c) for any best response weight assignment $w_p^*$ for any player $p$, $\sum_e w_p^*(e) u_p(e) - \sum_e w_p(e) u_p(e) \leq \epsilon$. (3a) is immediate, and we have $\sum_e w_e(p) = 1$. For (3b), recall that $y_p$ is the exact best response to $x_{-p}$. Therefore, for any player pair $p$ and $q$ and strategy $s$ of $q$, we could find a weight assignments $w$ on all edges $e$ (specifically, the weight assignments that made this a best response) such that $\sum_{e:s' \in e} w_p(e) = y_p^{s'}$ (for all $s'$, strategies of $p$), and $\sum_{e:s \in e} w_p(e) = x_q^s$. Since $w$ was a weight function for $y$, we also have $\sum_{e:s \in e} w_q(e) = y_q^s$. We are told that $|y_q^{*s} - x_q^s|$ is at most $\epsilon_1$, so we have $|\sum_{e:s \in e} w_p(e) - \sum_{e:s \in e} w_q(e)| \leq \epsilon_1$.

Condition (3c): As above, we can define the weight function $w_p^*(e)$ that makes $y_p$ a best response against $x_{-p}$. Also define any weight assignment $w_p$ that gives (for all strategies $s$ of $p$) $\sum_{e:s \in e} w_p(e) = x_p^s$. For any strategy $s$ of $p$, we are told that $|y_p^s - x_p^s| < \epsilon_1$, so we can say for any strategy $s$ of $p$, $\sum_{e:s \in e} w_p^*(e) - \sum_{e:s \in e} w_p(e) < \epsilon_1 \Rightarrow \sum_{e:s \in e} w_p^*(e) u_p(e) - \sum_{e:s \in e} w_p(e) u_p(e) < \epsilon_1 U_{\max}$. This means $\sum_s |\sum_{e:s \in e} w_p^*(e) u_p(e) - \sum_{e:s \in e} w_p(e) u_p(e)| < n\epsilon_1 U_{\max}$ if $n$ is the number of strategies for $p$. We can remove the absolute values because $w^*$ was a best response, giving $\sum_e w_p^*(e) u_p(e) - \sum_e w_p(e) u_p(e) \leq \epsilon$ as required, as long as $\epsilon_l \leq \epsilon \left(\frac{1}{nU_{\max}}\right)$.

$\square$

**Lemma 4.14.** *Given a $q \times r$ matrix $A$, an $q$-vector $b$ such that $Ax \geq b$ is feasible, and $p \in \mathbb{R}^r$ and $e \in \mathbb{R}^q$ such that $Ap \geq b - e$, there exists $p'$ satisfying $Ap' \geq b$ such that $\|p - p'\|_\infty$ is at most $e_{max}(A_{max} D_{max})^q$, where $e_{max} = \max_j e_j$, $A_{max}$ equals $\max_{ij} |A_{ij}|$ and $D_{max}$ is the largest determinant, in absolute value, of any submatrix of the matrix consisting of the columns of $A$ and $b$.*

*Proof.* We find a point $y$ such that $Ay \geq b - Ap$ and $\|y\|_\infty \leq e_{\max}(A_{\max} D_{\max})^q$. Setting $p' = p + y$ gives us the desired lemma. We first note that $Ay \geq b - Ap$ is feasible since it is satisfied by the point $x - p$, where $x$ satisfies $Ax \geq b$. Let $d$ equal $b - Ap$. So our goal is to find a $y$ satisfying $Ay \geq d$. By our assumption on $p$, we have $d_i \leq e_{\max}$ for all $i$.

Consider the following algorithm for constructing $y$. Set $y = 0$ and $L$ to be the empty LP. At the end of $i$ iterations, we will maintain the invariant that $\|y\|_\infty \leq (A_{\max} D_{\max})^i e_{\max}$. Find any constraint $A_k y \geq d_k$ not in $L$ that is not satisfied. (If no such constraint exists, then we are done.) Add this constraint to $L$. By the invariant on $|y|$, it follows that $|A_k y|$ is at most $A_{\max}^{i+1} D_{\max}^i e_{\max}$.

Since $d_k \leq e_{\max}$, it follows that $|d_k|$ is at most $A_{\max}^{i+1} D_{\max}^i e_{\max}$. Since the right hand side of every inequality of $L$ is at most this number, and the left hand side is a submatrix of $A$, by Cramer's rule there exists a vertex of $L$, every coordinate of which has magnitude at most $A_{\max}^{i+1} D_{\max}^i e_{\max}$ times the largest entry in the determinant of any submatrix of $A$, which is at most $D_{\max}$. (Note that $L$ is feasible since $Ay \geq d$ is feasible.) This yields the desired invariant $\|y\|_\infty \leq (A_{\max} D_{\max})^{(i+1)} e_{\max}$.

The above procedure stops in at most $q$ iterations, and yields a point $y$ such that $\|y\|_\infty \leq e_{\max}(A_{\max} D_{\max})^q$, thus completing the proof of the lemma. $\square$

**Corollary 4.15.** *Let $A$ be an $q \times r$ matrix, $b$ be an $q$-vector, and $p$ be the lexicographically smallest vector in $Ax \geq b$. Let $b' \in \mathbb{R}^q$ be such that $Ax \geq b'$ is feasible. If $p'$ is the lexicographically smallest vector in $Ax \geq b'$, then $\|p - p'\|_\infty$ is at most $e_{max}(A_{max} D_{max})^{qr+r(r+1)}$, where $e_{max} = \max_j |b_j - b'_j|$, $A_{max} = \max_{i,j} |A_{ij}|$, and $D_{max}$ is the largest determinant, in absolute value, of any submatrix of the matrix consisting of columns from $A$ and $b$.*

*Proof.* Let $L$ denote the LP $Ax \geq b'$. We apply Lemma 4.14 with $(A, p, b, e)$ replaced by $(A, p, b', b - b')$ to obtain a point $p'$ satisfying $L$ such that $|p_1 - p'_1|$ is at most $e_{\max}(A_{\max} D_{\max})^q$. We add a constraint $x_1 = p'_1$ to the LP $L$ and apply Lemma 4.14 with $(A, p, b, e)$ replaced by $(\tilde{A}, p, \tilde{b}, \tilde{e})$, where $\tilde{A}$ is the constraint matrix of $L$, $\tilde{b}$ is the right-hand side of $L$, and $\tilde{e}$ is the vector obtained by adding two additional coordinates to $e$, each with magnitude at most $|p_1 - p'_1|$ (for the two new inequality constraints resulting from the addition of $x_1 = q_1$). We obtain a new point $p'$ satisfying $L$ such that $|p_1 - p'_1|$ and $|p_2 - p'_2|$ are both at most $e_{\max}(A_{\max} D_{\max})^{2q+2}$. Repeating this for all the coordinates yields the lexicographically smallest vector $p'$ of $Ax \geq b'$ with $\|p - p'\|_\infty$ at most $e_{\max}(A_{\max} D_{\max})^{qr+r(r+1)}$. $\square$

## 5 Scarf's Lemma and Fractional Stability Problems

This section discusses the complexity of a number of well-known combinatorial problems that can be categorized as *fractional stability problems*. We begin with Scarf's Lemma, a fundamental result in combinatorics, originally introduced to prove that every balanced cooperative game with non-transferable utilities has a nonempty core (see Section 5.3) [41]. The core (no pun intended) of Scarf's proof is an elegant and constructive combinatorial argument, which has been applied to diverse combinatorial problems, including fractional stable matchings in hypergraphic preference systems, strong kernels in digraphs, and the fractional stable paths problem [2, 1, 28, 21]. We first show that the computational version of Scarf's lemma is **PPAD**-complete. We then establish the **PPAD**-completeness of stable matchings in hypergraphic preference systems, strong kernels in digraphs, core of balanced games with non-transferable utility, the fractional stable paths problem, and a fractional version of the Bounded Budget Connection game [30, 31].

### 5.1 Scarf's Lemma

In the computational version of Scarf's lemma (SCARF) we are given matrices $B, C$ and a vector $b$ satisfying the conditions in Theorem 5.1, and the goal is to find $\alpha \in \mathbb{R}_+^n$ satisfying the desired properties.

**Theorem 5.1.** *(Scarf's lemma [41]) Let $I = [\delta_{ij}]$ be an $m \times m$ identity matrix. Let $[n] = \{1, 2, \ldots, n\}$. Let $m < n$ and let $B$ be an $m \times n$ real matrix such that $b_{ij} = \delta_{ij}$ for $1 \leq i, j \leq m$. Let $b$ be a non-negative vector in $\mathbb{R}^m$, such that the set $\{\alpha \in \mathbb{R}_+^n : B\alpha = b\}$ is bounded. Let $C$ be an $m \times n$ matrix such that $c_{ii} \leq c_{ik} \leq c_{ij}$ whenever $i, j \leq m$, $i \neq j$ and $k > m$. Then there exists a subset $J \subset [n]$ of size $m$ such that*

**(P1)** $B\alpha = b$ for some $\alpha \in \mathbb{R}^n_+$ such that $\alpha_j = 0$ whenever $j \notin J$, and

**(P2)** For every $k \in [n]$ there exists $i \in [m]$ such that $c_{ik} \leq c_{ij}$ for all $j \in J$.

A subset $J \subset [n]$ of size $m$ is called a *feasible basis* of $(B, b)$ if it satisfies **(P1)**, and *subordinating* if it satisfies **(P2)**. To compute $\alpha$ of SCARF, it suffices to have a $J \subseteq [n]$ that is simultaneously subordinating and a feasible basis. Once such $J$ is computed, $\alpha$ can be computed by solving a system of linear equations. Also, given a solution $\alpha$, $J$ is easy to compute, since $J$ is $\alpha$'s support. Hence finding $\alpha$ and $J$ are computationally equivalent, to within polynomial time. In Theorem 5.2, we argue that Scarf's original proof [41], together with Todd's orientation technique [46], gives an end of the line argument for the existence of a subordinating and feasible basis, thus showing that SCARF is in **PPAD**.

> SCARF: Given matrices $B$ and $C$ and vector $b \in \mathbb{R}^m$, all obeying the requirements from Theorem 5.1, find a subset of $m$ column indices that is a feasible basis for $(B, b)$ and is subordinating for $C$.

**Theorem 5.2.** SCARF $\leq_P$ END OF THE LINE.

*Proof.* The pair $(B, b)$ is *non-degenerate* if $b$ is not in the cone spanned by fewer than $m$ columns of $B$. We call $C$ ordinal-generic if all the elements in each row of $C$ are distinct. There exists a small perturbation $b'$ of $b$ such that the pair $(B, b)$ is non-degenerate and every feasible basis for $(B, b')$ is also a feasible basis for $(B, b)$. By slightly perturbing $C$, we can obtain an ordinal-generic matrix $C'$ satisfying the assumptions of the theorem, and if the perturbation is small enough, then any subordinating set for $C'$ is also subordinating for $C$. Hence, we may assume that $(B, b)$ is non-degenerate, and that $C$ is ordinal-generic.

Lemma 5.3 is well-known. Its proof requires that $\{\alpha \in \mathbb{R}^n_+ : B\alpha = b\}$ is bounded and $(B, b)$ is non-degenerate. For the proof of Lemma 5.4, we refer the reader to [41] or [2] or page 1127 of Schrijver's Combinatorial Optimization book [42]. Proof of Lemma 5.4 uses the assumption that $C$ is ordinal-generic.

**Lemma 5.3.** *Let $J$ be a feasible basis for $(B, b)$, and $k \in [n] \setminus J$. Then there exists a unique $j \in J$ such that $J + k - j$ (i.e., $J \cup \{k\} \setminus \{j\}$) is a feasible basis. Also, given $J$ and $k$, we can find $j$ in polynomial time.*

**Lemma 5.4.** *(Scarf [41]) Let $K$ be a subordinating set for $C$ of size m-1. Then there are precisely two elements $j \in [n] \setminus K$ such that $K + j$ is subordinating for $C$, unless $K \subseteq [m]$, in which case there exists precisely one such $j$. Given $K$, we can find values of $j$ in polynomial time.*

The natural pivot rules arising from Lemma 5.3 and Lemma 5.4 are called the *feasible pivot rule* and the *ordinal pivot rule* respectively.

The original proof of Scarf's lemma ([41], [2]) uses an "*undirected* end of the line argument", thus showing its **PPA**-membership. It is easy to see that **PPAD** $\subseteq$ **PPA**, however it is unknown if **PPAD** = **PPA**. To prove **PPAD**-membership of SCARF, we need a "*directed* end of the line argument". Shapley [44] presented a geometric orientation rule for the equilibrium points of (non-degenerate) bimatrix games based on the Lemke-Howson algorithm [32]. Extending Shapley's rule, Todd [46] developed a similar orientation theory for generalized complementary pivot algorithms. We now apply Todd's orientation technique to prove **PPAD**-membership of SCARF.

Let $X = \{1, 2, \ldots, n\}$. A subset of $X$ of cardinality $m$ is called an $m$-subset. Let $X_m$ denote the collection of ordered (with the natural ordering defined by $X$) $m$-tuples of distinct elements of $X$. Two $m$-tuples in $X_m$ are *equivalent* iff one is an even permutation of the other. Let $P$ be any element of an equivalent set. We denote the corresponding equivalent set by $\overline{P}$. If $P' \in X_m$ is an odd permutation of $P \in X_m$, then we call $\overline{P'}$ the *negative* of $P$ and write $\overline{P'} = -\overline{P}$. Let $P = (e_1, \ldots, e_n) \in X_n$. For $\mu = \pm 1$, we say $\overline{P}$ contains $\mu(\overline{P \backslash e_i})$ positively (negatively) if $\mu(-1)^i$ is positive (negative).

Let $e \in X$ be a specific element. Let $\mathcal{F}$ be the set of all feasible bases containing $e$, and $\mathcal{S}$ be the set of all subordinating sets of size $m$ *not* containing $e$. Note that both $\mathcal{F}$ and $\mathcal{S}$ are $m$-subsets of $[n]$. Let $V(\mathcal{F}, \mathcal{S}, e)$ be the set of pairs $(\overline{F}, \overline{S}) \in \mathcal{F} \times \mathcal{S}$ satisfying either (i) $\overline{F} = \pm\overline{S}$ (called a matched pair) or (ii) $e \in F, e \notin S$ and $F \backslash S = \{e\}$ (called an unmatched pair). A matched pair $(\overline{T}, \overline{T})$ is positive, while $(\overline{T}, -\overline{T})$ is negative. An unmatched pair $(\overline{F}, \overline{S})$ is positive (negative) if $\overline{F}$ is contained in $(\overline{S \cup e})$ positively (negatively).

**Lemma 5.5.** *(Todd [46]) (a) Every matched pair is adjacent to one unmatched pair by a feasible pivot, or one unmatched pair by a ordinal pivot, but not both. (b) Every unmatched pair is adjacent to one pair by a feasible pivot and one pair by a ordinal pivot.*

**Lemma 5.6.** *(Todd [46]) (a) If two unmatched pairs are adjacent by a feasible pivot, they have opposite signs. (b) If a matched pair and an unmatched pair are adjacent by a feasible pivot, they have the same sign. (c) If two pairs are adjacent by a ordinal pivot, they have opposite signs.*

Similar to [46], we construct a directed graph with vertices representing the pairs in $V(\mathcal{F}, \mathcal{S}, e)$. If two unmatched pairs are adjacent by a feasible pivot, we add a directed edge from the negative pair to the positive pair. If a matched pair is adjacent by a feasible pivot to an unmatched pair, we add a directed edge from the matched pair to the unmatched pair if both are positive and in the reverse direction if both are negative. If two pairs are adjacent by an ordinal pivot, we add a directed edge from the positive pair to the negative pair. From Lemmas 5.5 and 5.6, it follows that each unmatched pair has indegree 1 and outdegree 1. Each matched pair has indegree 0 and outdegree 1 if positive, and indegree 1 and outdegree 0 if negative. It is easy to see that $[m]$ is in $\mathcal{F}$ and is not subordinating. By Lemma 5.4 there exists $f \neq e$ such that $[m] - e + f$ is in $\mathcal{S}$. We shall use the pair $([m], [m] - e + f)$ as the initial source of END OF THE LINE. This gives the required **PPAD** property. □

In Section 5.2, we establish the **PPAD**-hardness of FRACTIONAL HYPERGRAPH MATCHING, which reduces to SCARF in polynomial time [1], thus completing the proof that SCARF is **PPAD**-complete.

### 5.2 Hypergraphic Preference Systems

A hypergraphic preference system is a pair $(H, \mathcal{O})$, where $H = (V, E)$ is a hypergraph, and $\mathcal{O} = \{\preceq_v : v \in V\}$ is a family of linear orders, $\preceq_v$ being an order on the set of edges containing the vertex $v$. A set $M$ of edges is called a *stable matching* with respect to the preference system if (a) it is a matching and (b) for every edge $e$ there exists a vertex $v \in e$ and an edge $m \in M$ containing $v$ such that $e \preceq_v m$. A nonnegative function $w$ on the edges in $H$ is called a *fractional matching* if $\sum_{v \in h} w(h) \leq 1$ for every vertex $v$. A fractional matching $w$ is called *stable* if every edge $e$ contains a vertex $v$ such that $\sum_{v \in h, e \preceq_v h} w(h) = 1$.

Aharoni and Fleiner [1] used Scarf's lemma to prove that every hypergraphic preference system has a fractional stable matching. This naturally leads to a computational problem – FRACTIONAL

HYPERGRAPH MATCHING : given a hypergraphic preference system $(H, \mathcal{O})$, find a fractional stable matching.

> FRACTIONAL HYPERGRAPH MATCHING: Given a hypergraph $H$ and a preference ordering $\mathcal{O}$ for each node in $H$, find a weight assignment across the edges such that the weight adjacent to each vertex is at most 1 and each edge $e$ includes some vertex $v$ that is adjacent to weight exactly 1 on edges preferred by $v$ at least as much as $e$.

We first observe that the proof of [1] is a polynomial time reduction from FRACTIONAL HYPERGRAPH MATCHING to SCARF, thus placing it in **PPAD**. We now show that FRACTIONAL HYPERGRAPH MATCHING is **PPAD**-hard via a reduction from preference games.

**Theorem 5.7.** DEGREE $d$ PREFERENCE GAME $\leq_P$ FRACTIONAL HYPERGRAPH MATCHING.

*Proof.* We are given a preference game over players $[n] = \{1, \ldots, n\}$. We construct the following hypergraph matching instance $(H, \mathcal{O})$, $H = (V, E)$. The set $V$ of vertices is $[n] \cup \{i^* : i \in [n]\}$; that is, we have two vertices $i$ and $i^*$ for each player $i$. The set of edges is given by the following.

$$\{\{i^*\} : i \in [n]\} \bigcup \{\{i, i^*\} \cup J_i : i \in [n], J_i \subseteq \mathrm{in}(i)\}$$

(Note that $J_i$ is a subset of players that prefer $i$ over themselves.)

We next describe the linear order for a given vertex $i$. Let $e_1$ and $e_2$ be two edges containing $i$. By our construction of $E$, there exists a unique $i_1$ such that $\{i_1, i_1^*\}$ is a subset of $e_1$. Similarly, there is a unique $i_2$ such that $\{i_2, i_2^*\}$ is a subset of $e_2$. If $i_1 \neq i_2$, then we require that $e_1 \succeq_i e_2$ if and only if $i_1 \succeq_i i_2$. If $i_1 = i_2$, then we require the following condition on $\succeq_i$: $e_1 \succeq_i e_2$ whenever $e_1 \supseteq e_2$. Finally, for any vertex $i^*$, we select any linear order in which $e_1 \succeq_{i^*} e_2$ whenever $\{i, i^*\}$ is a subset of $e_1$ and $e \succeq_{i^*} \{i*\}$ for all $e$.

The number of vertices in the above hypergraph is $2n$, and the number of edges is at most $n(2^d + 1)$, where $d$ is the maximum in-degree of the preference game. Since we are given a preference game of constant degree, the above construction is polynomial time.

We show that there is a stable solution for the preference game if and only if there is a stable fractional matching for the hypergraph preference system. Suppose $w$ is a stable solution for the preference game: $w_{ij}$ represents the weight assigned by player $i$ to player $j$. For a given player $j$, we sort all the players $i$ in $\mathrm{in}(j)$ in nonincreasing order of the $w_{ij}$ values; let the order be $j_1, j_2, \ldots, j_{d_j}$, where $d_j$ is the in-degree of $j$. To every edge of the form $\{j, j^*\} \cup \{j_1, \ldots, j_k\}$, $1 \leq k < d_j$, we assign the weight $w_{j_k j} - w_{j_{k+1} j}$. We assign weight $w_{j_{d_j}}$ to the edge $\{j, j^*\} \cup \mathrm{in}(j)$ and weight $w_{jj} - w_{j_1 j}$ to the edge $\{j, j^*\}$. Finally, we assign weight $1 - w_{jj}$ to the edge $\{j^*\}$. This ensures the following:

$$\begin{aligned} \sum_{e : \{j, j^*\} \in e} f(e) &= w_{jj} \quad \text{for all } j \\ \sum_{e : \{j, j^*, i\} \in e} f(e) &= w_{ij} \quad \text{for all } j, i \in \mathrm{in}(j) \end{aligned}$$

We next argue that the fractional matching $f$ thus defined is stable.

There are three types of edges for us to consider. (1) $e = \{j, j^*, j_1, j_2, \ldots, j_k\}$ for some $j, k$, (2) $e = \{j, j^*\} \cup S$ for some $j, S \neq \{j_1, j_2, \ldots, j_k\}$ for any $k$, and (3) $e = \{j^*\}$ for some $j$

First consider $e = \{j, j^*, j_1, j_2, \ldots, j_k\}$ for some $j, k$. We separate this into two cases. The first case is when there is no proper subset of $e$ that has positive weight. In this case, we argue that $j$ is a vertex in $e$ such that $\sum_{h \succeq_j e} f(h)$ equals 1. $\sum_{h \succeq_j e} f(h) = \sum_{i \succeq_j j} \sum_{h : \{i, i^*, j\} \in h} f(h) + \sum_{e \subseteq h} f(h) =$ $\sum_{i \succeq_j j} w_{ji} + \sum_{h:e \subseteq h} f(h) + \sum_{h:h \subset e} f(h) = \sum_{i \succeq_j j} w_{ji} + \sum_{h:\{j,j^*\}h} f(h) = \sum_{i \succeq_j j} w_{ji} + w_{jj} = 1$.
The second case is when there is some proper subset $e'$ of $e$ with positive weight. Say $e' = \{j, j^*, j_1, j_2, \ldots, j_{s-1}\}$ for $s \leq k$. Because $s \leq k$, $j_s \in e$. We will show that $j_s$ is a vertex in $e$ such that $\sum_{h \succeq_{j_s} e} f(h)$ equals 1. Since $e'$ has positive weight, $w_{j_{s-1}j} - w_{j_s j} > 0 \Rightarrow w_{j_s j} < w_{jj}$. Therefore, since $w$ was a preference game equilibrium, $\sum_{i \succeq_{j_s} j} w_{j_s i} = 1$. So, $\sum_{i \succeq_{j_s} j} \sum_{h:\{i,i^*,j_s\} \in h} f(h) = 1 \Rightarrow \sum_{h:\{j,j^*,j_s\} \in h} f(h) + \sum_{i \succeq_{j_s} j, i \neq j} \sum_{h:\{i,i^*,j_s\} \in h} f(h) = 1 \Rightarrow \sum_{h \succeq_{j_s} e} f(h) = 1$.

Next consider $e = \{j, j^*\} \cup S$ for some $j, S \neq \{j_1, j_2, \ldots, j_k\}$ for any $k$. Now, pick edge $e' \supset e$, $e' = \{j_1, j_2, \ldots, j_k\}$ for $j_k \in e$. Again, we can separate this into two cases based on whether or not there is a proper subset of $e'$ with positive weight. If there is no such proper subset, then $j$ will have $\sum_{h \succeq_j e} f(h) = 1$, but the same argument as above. If there is a proper subset $e'' \subset e'$ with positive weight, we will argue that $j_k$ satisfies $\sum_{h \succeq_{j_k} e} f(h) = 1$. Since $\{j, j^*\} \in e''$, $j_k \notin e''$, $\sum_{h=\{j,j^*\} \cup S} f(h) \geq f(e'') + \sum_{h:\{j,j^*,j_k\} \in h} f(h) \Rightarrow w_{jj} \geq f(e'') + w_{j_k j}$. We picked $e''$ such that $f(e'') > 0$, so $w_{jj} > w_{j_k j}$. As in the previous paragraph, this implies that $\sum_{h \succeq_{j_k} e} f(h) = 1$.

To complete this direction of the lemma, consider an edge $\{j^*\}$ for some $j$. By our construction, this is the least preferred edge for $j^*$, and the assignment of weight $1 - w_{jj}$ guarantees that the sum of the weights of all edges containing $j^*$ equals 1.

We next prove the other direction of the lemma. Suppose $f$ is a stable fractional matching for the hypergraph preference system. We construct the following assignment for the preference game. We set $w_{ij}$ to be the sum of the weights of edges containing the subset $\{j, j^*, i\}$. It is easy to see that $w_{ij} \leq w_{jj}$ for all $i$ and $j$. It remains to argue the stability of $w$.

We first claim that if $f$ is stable, then for any $S_1$ and $S_2$ such that $S_1 - S_2$ and $S_2 - S_1$ are both nonempty, at most one of $f(\{j, j^*\} \cup S_1)$ and $f(\{j, j^*\} \cup S_2)$ is positive. To see this, observe that if both are positive, then for every vertex $v$ in the edge $e = \{j, j^*\} \cup S_1 \cup S_2$, the sum of weights assigned to edges that are at least as much preferred by $v$ as $e$ is less than one since $v$ is in either $\{j, j^*\} \cup S_1$ or $\{j, j^*\} \cup S_2$, both of which have positive weight and are less preferred than $e$ by $v$. This implies that such a matching could not be stable for edge $e$. Thus, in $f$, the positive weights to edges containing $\{j, j^*\}$ are all assigned to a chain of edges $e_1 \subset e_2 \ldots \subset e_k$, for some $k$. Define $e_{k+1}$ to be $\{j, j^*\} \cup \text{in}(j)$. We next observe that for every vertex $v$ in $e_i - e_{i-1}$, $1 < i \leq k+1$, the sum of the weights of the edges $v$ prefers at least as much as $e_i$ equals 1. This is because such a vertex exists in $e_{i-1} \cup \{v\}$ (by the definition of stable matching) and $v$ is the only possibility.

Consider any $w_{i\ell} > 0$. To establish stability of $w$, we prove by a contradiction that for all $j$ such that $j \succeq_i \ell$, $w_{ij} = w_{jj}$. Suppose not, then there exists a $j$ such that $j \succeq_i \ell$, and two edges $e, e' \supseteq \{j, j^*\}$ with $i \in e$, $i \notin e'$, and $f(e') > 0$. Let $e$ denote the smallest edge containing $i$ in the chain $e_1 \subset e_2 \ldots \subset e_{k+1}$ mentioned in the preceding paragraph. (Since $i \in e_{k+1}$, $e$ exists.) By the argument above, the sum of the weights of the edges $i$ prefers at least as much as $e$ equals 1. However, $w_{i\ell} > 0$ implies that there exists an edge $e'' \supseteq \{\ell, \ell^*, i\}$ with $f(e'') > 0$, yielding a contradiction since $i$ prefers $e$ over $e''$. $\square$

### 5.3 Cooperative Games with Non-Transferable Utilities

**Definition 5.8.** *A game with non-transferable utilities over $n$ players is specified by a function $V$ that for each subset $S$ of $N = \{1, 2, \ldots, n\}$ returns a set $V(S)$ of outcomes – each outcome being a vector of utility values, one component for each player in $S$. A collection $T$ of coalitions is* balanced *if there exists an assignment of reals $\delta_S$ for each coalition $S$ in $T$ such that for all $v$, $\sum_{S:v \in S} \delta_S = 1$. We say that $u$ is* attainable *by $S$ if $u \in V(S)$. A game is* balanced *if and only if for any balanced* *collection $T$ and any $u$, if $u_S$ is attainable by all $S$ in $T$, then $u$ is attainable by $N$.*

As mentioned earlier, Scarf [41] proved that every balanced game has a nonempty core. We define CORE-BALANCED-NTU below, a natural computational version of this claim. Scarf's proof [41], which is a reduction to SCARF, together with Theorem 5.9 establish its **PPAD**-completeness.

> CORE-BALANCED-NTU: The game is specified by a set $N$ of players, a collection $\mathcal{S}$ of proper subsets of $N$ (the coalitions), and for each $S \in \mathcal{S}$, vectors $u_1, \ldots, u_{k_S}$ in $\mathbb{R}^{|S|}$ such that $V(S) = \{u \in \mathbb{R}^{|S|} : \exists j \ u \leq u_j\}$. For a coalition $S \notin \mathcal{S}$, $V(S) = \{0\}^{|S|}$ and $V(N)$ is defined as the set of all $u$ for which there exists a balanced collection $T$ such that $u_S$ is attainable by all $S$ in $T$. The goal is to find an element in the core.

**Theorem 5.9.** FRACTIONAL HYPERGRAPH MATCHING $\leq_P$ CORE-BALANCED-NTU.

*Proof.* Suppose we are given a hypergraph $H$ and for each vertex $i$, a preference ranking among all edges containing $i$. We first add, for each vertex $i$ in $H$, a new vertex $i^*$ and edge $\{i, i^*\}$. We set the preference of $i$ for the edge $\{i, i^*\}$ to be the least among all the edges containing $i$. Let $N$ denote the new set of nodes and $E$ the new set of edges. For $S \in E$ and $i \in N$, let $r_i(S)$ denote the rank of $S$ in $i$'s preference list, with 0 assigned to the least preferred edge (thus for every $i$, $r_i(\{i, i^*\}) = 0$). We now define a balanced cooperative game with non-transferable utilities. For each node in $N$, we have a player in the game. For any coalition $S$, we consider two cases. If $S \in E$, then we have a single vector $r_S = (r_{i_1}(S), r_{i_2}(S), \ldots, r_{i_{|S|}}(S))$, where $S = \{i_1, i_2, \ldots, i_{|S|}\}$. Note that by definition, if $S \notin E$ and $S \neq N$, then $V(S)$ equals $0^{|S|}$.

For $N$, note that $V(N)$ is precisely the set of all $u$ such that $u_S$ is attainable by all $S$ in some balanced collection $T$. We first observe that we can determine in polynomial time whether a given $u$ is in $V(N)$. For each $S$, if $u \leq r_S$, then we have a variable $x_S$ for $S$. Now we simply solve the linear program:

$$\sum_{S: i \in S} x_S = 1$$

It is easy to see that the linear program is feasible if and only if $u$ is in $V(N)$. Consider any balanced collection $T$; if we have a $u$ such that $u_S$ is attainable by all coalitions $S$ in $T$, then the above linear program would be feasible – the $\delta_S$ values that verify the balanced condition yield the solution for the above LP, and hence $u$ is attainable by $N$. For the other direction, consider any $u$ that is attainable by $N$. Then, by our construction the above linear program is feasible. The $x_S$ values we obtain precisely specify the $\delta_S$ values, meaning that $u_S$ is attainable by every $S$ for which $\delta_S > 0$.

It is straightforward to compute the above reduction in time polynomial in $H$. We finally claim that from an element of the core, a fractional stable hypergraph matching can be obtained in polynomial time. Suppose $u$ is in the core. Since $u$ is attainable by $N$, we find the $x_S$ that satisfy the above linear program. We claim that the weights $x_S$ yield a stable fractional hypergraph matching in $H$. Consider any edge $S'$. Since $u$ is in the core, there exists a player $i$ in $S'$ such that the utility for $i$ in $u$ is at least as high as that for $i$ in $V(S')$. Since $u$ is attained by $N$, the utility (preference) of $i$ in each $S$ for which $x_S > 0$ is also at least as high as that of $i$ in $S'$. Thus, $x_S$ yields a stable matching. □

## 5.4 Fractional Stable Paths Problem

The Fractional Stable Paths problem, introduced in [21], is defined as follows. Let $G$ be a graph with a distinguished destination node $d$. Each node $v \neq d$ has a list $\pi(v)$ of simple paths from $v$ to $d$ and a preference relation $\succeq_v$ among the paths in $\pi(v)$. For a path $S$, we also define $\pi(v, S)$ to be the set of paths in $\pi(v)$ that have $S$ as a suffix. A *proper suffix* $S$ of $P$ is a suffix of $P$ such that $S \neq P$ and $S \neq \emptyset$.

A *feasible fractional paths solution* is a set $w = \{w_v : v \neq d\}$ of assignments $w_v : \pi(v) \to [0, 1]$ satisfying:

1. **Unity condition**: for each node $v$, $\sum_{P \in \pi(v)} w_v(P) \leq 1$

2. **Tree condition**: for each node $v$, and each path $S$ with start node $u$, $\sum_{P \in \pi(v,S)} w_v(P) \leq w_u(S)$.

In other words, a feasible solution is one in which each node chooses at most 1 unit of flow to $d$ such that no suffix is filled by more than the amount of flow placed on that suffix by its starting node. A feasible solution $w$ is *stable* if for any node $v$ and path $Q$ starting at $v$, one of the following holds:

**(S1)** $\sum_{P \in \pi(v)} w_v(P) = 1$, and for each $P$ in $\pi(v)$ with $w_v(P) > 0$, $P \geq_v Q$; or

**(S2)** There exists a proper suffix $S$ of $Q$ such that $\sum_{P \in \pi(v,S)} w_v(P) = w_u(S)$, where $u$ is the start node of $S$, and for each $P \in \pi(v, S)$ with $w_v(P) > 0$, $P \geq_v Q$.

In other words, in a stable solution: if node $v$ has not fully chosen paths that it prefers at least as much as $Q$, then it has completely filled path $Q$ by filling some suffix with paths it prefers at least as much as $Q$.

We define a computational version, FRACTIONAL SPP.

> FRACTIONAL SPP: Given a graph, a destination node, and a preference list for each node across paths to the destination. Find a weight assignment for each node to the paths in its preference list that is both feasible (satisfies the Unity and Tree conditions) and is stable (every path satisfies property (S1) or (S2)).

We note that Haxell and Wilfong [22] show FRACTIONAL HYPERGRAPH MATCHING $\leq_P$ FRACTIONAL SPP (see Section 5.2), and the problem of finding a fractional co-acyclic kernel (related to STRONG KERNEL, see Section 5.5) can also be reduced to FRACTIONAL SPP. Together with our reduction from FRACTIONAL SPP to PERSONALIZED EQUILIBRIUM (see Theorem 5.11 below), this gives an alternate proof of **PPAD**-membership for these two problems.

### 5.4.1 PPAD-completeness
**Theorem 5.10.** PREFERENCE GAME $\leq_P$ FRACTIONAL SPP.

*Proof.* We are given a preference game over player set $[n]$, including preference relation $\succeq_i$ for all $i \in 1 \ldots n$. We will convert this into a fractional stable paths problem. Create a node $v_i$ for each $i$. Also create a universal destination node $d$. For all $i$, define $P_{ii}$ = the path $(v_i, d)$. For all $i, j$, define $P_{ij}$ = the path $(v_i, v_j, d)$. Let $\pi(v_i)$ (the set of preferred paths for $v_i$) = $\{P_{ij} : j \succeq_i i\}$. If $k \succeq_i j$, then $P_{ik} \succeq_i P_{ij}$.

Let $w_i(j)$ refer to the amount of weight placed by node $v_i$ on path $P_{ij}$ in a fractional SPP solution, and let $w_i(i)$ be the amount of weight placed by $i$ on path $P_{ii}$. Now we will show that $w$ is a fractional stable paths solution if and only if $w$ defines an equilibrium of the preference game.

**$w$ is a fractional stable paths solution $\Rightarrow$ $w$ is an equilibrium of the preference game.**.
By the unity condition, for each $i$, $\sum_{j: P_{ij} \in \pi(v_i)} w_i(j) \leq 1 \Rightarrow \sum_j w_i(j) \leq 1$. $P_{ii}$ starts at $v_i$, and there is no proper final suffix of $P_{ii}$, so condition (S1) must apply for $P_{ii}$. Therefore, $\sum_{j: P_{ij} \in \pi(v_i)} w_i(j) = \sum_j w_i(j) = 1$, as required for the preference game. By the tree condition, for any $i, j$, $\sum_{P \in \pi(v, P_{jj})}$ weight on $P \leq w_j(j) \Rightarrow w_i(j) \leq w_j(j)$. So $w$ is a feasible weight assignment for the preference game.

Now suppose for contradiction that $w$ is not lexicographically maximal (with respect to $w_{-i}$) for player $i$ in the preference game. Then, there is some feasible weight assignment $w'$ and some $j$ such that $\sum_{k \succeq_i j} w_i(k) < \sum_{k \succeq_i j} w'_i(k)$. Take the lexicographically maximal such $w'$ and the highest preference such $j$ (from $i$'s preference list). By this choice of $w'$ and $j$, $\sum_{k \succ_i j} w_i(k) = \sum_{k \succ_i j} w'_i(k)$, so $\sum_{k =_i j} w_i(k) < \sum_{k =_i j} w'_i(k)$. There must be some $j'$ with $j' =_i j$ such that $w_i(j') < w'_i(j')$. Consider path $P_{ij'}$. (S2) is not true by our choice of $j'$ and the fact that $w'$ was a feasible solution (so $w'_i(j') \leq w_{j'}(j')$). However, since $\sum_{k =_i j'} w_i(k) < \sum_{k =_i j'} w'_i(k)$, there must be some path $P_{ik}$ such that $k \prec_i j'$ with $w_i(k) > 0$. So (S1) is also not true, and $w$ was not a stable solution - a contradiction.

**$w$ is an equilibrium of the preference game $\Rightarrow$ $w$ is a fractional stable paths solution.**.
We know that $\sum_j w_i(j) = 1$ for all $i$. This immediately satisfies the unity condition. Since $w$ is a feasible set of weights for the preference game, $w_i(j) \leq w_j(j)$ for all $i, j$. This means that the weight placed on $P_{ij}$ is at most the weight placed on $P_{jj}$. Since $P_{ij}$ is the only path from $v_i$ that passes through node $v_j$, the tree condition holds. Now consider any path $P_{ij}$ from node $i$. Case 1: $w_i(j) = w_j(j)$. In this case, condition (S2) is satisfied. Case 2: $w_i(j) < w_j(j)$. Because $w$ was lexicographically maximal, any weight assignment $w'$ with $\sum_{k \succeq_i j} w'_i(k) \geq \sum_{k \succeq_i j} w_i(k)$ must be infeasible. We said that $w_i(j) < w_j(j)$, so it is only possible for all such $w'$ to be infeasible if $\sum_{k \succeq_i j} w_i(k) = 1$. Then $\sum_{k \prec_i j} w_i(k) = 0$, so (S1) is satisfied. □

**Theorem 5.11.** FRACTIONAL SPP $\leq_P$ PERSONALIZED EQUILIBRIUM.

*Proof.* Suppose we are given an instance of FRACTIONAL SPP, consisting of a set of nodes $V$, a set of preferred paths $\pi(v)$ for all $v \in V$, and a preference relation $\succeq_v$ for each set $\pi(v)$. We can also find $\pi(v, S)$, the set of all $P \in \pi(v)$ such that $S$ is a subpath of $P$. Let $q_v(P) = $ the number of paths $Q$ such that $P \succeq_v Q$.

We will create the following instance of PERSONALIZED EQUILIBRIUM. The set of players is $V$. The set of strategies $S_v$ for a node $V$ is $\pi(v) \cup \{N\}$ ($N$ stands for "No path"). For node $v$, there is exactly one edge defined for each strategy. Edge $P'$ for strategy $P = \{S : P \in \pi(v, S)\}$. The edge for strategy $N$ ($N'$) is a singleton edge, containing only that strategy. The payoffs to player $v$ are: $u_v(P') = q_v(P) + 1$, $u_v(N) = 1$.

Suppose $w$ is a set of weights in a personalized equilibrium of the game defined above. $w_v(P')$ represents the weight assigned by $v$ to edge $P'$. We will show that $w$ is a personalized equilibrium if and only if $w' : w'_v(P) = w_v(P')$ is a fractionally stable solution to the FRACTIONAL SPP instance.

First, assume $w$ is a personalized equilibrium. Then, we know that for all $v$, the total weight placed by $v$ on all edges is 1, or $w_v(N) + \sum_{P \in \pi(v)} w_v(P') = 1$. Therefore, $\sum_{P \in \pi(v)} w_v(P') \leq$ $1 \Rightarrow \sum_{P \in \pi(v)} w'_v(P) \leq 1$, so the Unity condition holds. Also, the sum of weights placed by $v$ on edges adjacent to a strategy $S$ of another player $v'$ is at most $w_{v'}(S)$. That is, for path $S \in \pi(v')$ ($v' \neq v$), $\sum_{P':S \in P'} w_v(P') \leq w_{v'}(S') \Rightarrow \sum_{P \in \pi(v,S)} w_v(P') \leq w_{v'}(S') \Rightarrow \sum_{P \in \pi(v,S)} w'_v(P) \leq w'_{v'}(S)$, so the Tree condition holds. Finally, take any path $Q \in \pi(v)$. Case 1: The payoff to $v$ in the personalized equilibrium is at least $q_v(Q) + 1$. In this case, we know that $v$ puts a total of weight 1 on edges with payoff at least $q_v(Q) + 1$, or $\sum_{P':u_v(P') \geq q_v(Q)+1} w_v(P') = 1 \Rightarrow \sum_{P:P \succeq_v Q} w_v(P') = 1 \Rightarrow \sum_{P:P \succeq_v Q} w'_v(P) = 1$, so condition (S1) holds. Case 2: The payoff to $v$ in the personalized equilibrium is less than $q_v(Q) + 1$. Since this is a personalized equilibrium, it cannot be possible for $v$ to move some weight off a lower paying hyperedge onto a higher paying hyperedge. This includes moving weight from any of the edges in the equilibrium with payoff less than $q_v(Q) + 1$ to the edge $Q'$. By nature of the edges we've defined, if $P' \cap Q'$ for $P', Q' \in \pi(v)$, then either $P' \subset Q'$ or $Q' \subset P'$. This means that there is some $S \in Q'(S \in \pi(v')$ such that $\sum_{R':S \in R'} w_v(R') = w_{v'}(S')$ and for all $R' : S \in R'$, if $w_v(R') > 0$ then $q_v(R) \geq q_v(Q)$. So, $\sum_{R \in \pi(v,S)} w_v(R') = w_{v'}(S') \Rightarrow \sum_{R \in \pi(v,S)} w'_v(R) = w'_{v'}(S)$, and for all $R \in \pi(v,S)$ with $w'_v(R) > 0$, $R \succeq_v Q$, as required for condition (S2).

Next, assume $w'$ is a fractionally stable solution. We can assign weights $w_v(P') = w'_v(P)$, $w_v(N) = 1 - \sum_{P \in \pi(v)} w_v(P)$. The Unity condition ensures that $\sum_{P \in \pi(v)} w'_v(P) \leq 1$, $w_v(N) \geq 0$ and we have a set of weights that sum to 1 for any player $v$. The Tree condition says that $\sum_{P \in \pi(v,S)} w'_v(P) \leq w'_{v'}(S)$ for any $S \in \pi(v')$, which gives $\sum_{P':S \in P'} w_v(P') \leq w_{v'}(S')$, as required for a feasible solution. Finally, we must verify that $w_v$ is a best response for player $v$. Let $w^*$ be the best response weight function for $v$, and for the sake of contradiction, assume $w^*$ gives a better total payoff. Look at the edge $P'$ with the highest $q_v(P)$ such that $w^*_v(P') > w_v(P')$. By nature of the edges we've defined, if $P' \cap Q'$ for $P', Q' \in \pi(v)$, then either $P' \subset Q'$ or $Q' \subset P'$. Therefore, for all edges $P''$ with $q_v(P'') > q_v(P')$, if $w^*_v(P'') < w_v(P'')$, then we could increase $w^*_v(P'')$ and improve the payoff, so $P'$ is the highest utility edge in which $w_v$ and $w^*_v$ differ. Now look at edge $P'$ with weights $w$ in the fractional stable paths problem. Since $w^*_v(P') > w_v(P')$ and $w^*_v(P'') = w_v(P'')$ for all $P''$ with higher payoff than $P'$, then for all $S \in P'$ ($s \in \pi(v')$ for some $v'$), $\sum_{R':S \in R', q_v(R) > q_v(P)} w_v(R') < w_{v'}(S') \Rightarrow \sum_{R \in \pi(v,S), q_v(R) > q_v(P)} w'_v(R) < w'_{v'}(S)$, so condition (S2) is not satisfied. However, since $v$ puts less weight on edges with payoff at least as high as the payoff for $P'$, the total payoff to $v$ is $< q_v(P) + 1$. Therefore, $\sum_{R':q_v(R) \geq q_v(P)} w_v(R') < 1$, so $\sum_{R \succeq_v P} w'_v(R) < 1$, so condition (S1) is also not satisfied. This means that $w'$ was not a fractionally stable solution, contradicting our assumption. So $w$ must have been a best response weighting for each $v$. □

### 5.4.2 Special Cases of Fractional SPP

**Theorem 5.12.** *Fractional SPP is **PPAD**-hard even if each node's preference list consists of all paths, ordered shortest to longest based on edge length (where each node defines its own edge lengths, which may not obey triangle inequality).*

*Proof.* In the reduction from preference games to fractional SPP in Theorem 5.10, each path in any preference list has either 1 hop (a direct path to the destination $d$) or two hops. For each of these two hop paths $(i \to j \to d)$, let $Q_{ij}$ = the number of paths $P$ such that $P \geq_i (i \to j \to d)$. Notice that $(i \to j \to d) \geq_i (i \to k \to d)$ if and only if $Q_{ij} \leq Q_{ik}$. Now, define the following lengths $l$ on the edges of the graph from the perspective of node $i$. If $(i \to j \to d) \in \pi(i)$, then $l(i,j) = Q_{ij}$, $l(j,d) = 1$. $l(i,d) = Q_{ii} + 1$. Pick $M_i = \max_j Q_{ij} + 2$. Let $l(x,y) = M_i$ for all other edges. Now, any path $(i \to j \to d) \in \pi(i)$ will have length $Q_{ij}$, path $(i \to d)$ will have length $Q_{ii}$.

This preserves the preference list across these paths. Most other paths will have a last segment of length $M_i$, so will be longer than $l(i,d)$. The only exception is paths that pass through a $j$ such that $(i \to j \to d) \in \pi(i)$. However, for these paths, the only way to arrive at $j$ without following the direct edge $(i,j)$ would be to pass through an edge of length $M_i$, so these paths too will be longer than $l(i,d)$. □

**Theorem 5.13.** *Fractional SPP is **PPAD**-hard even if all preferred paths are preference-ordered based on the path length (where each node defines its own distances on the edge lengths, and these distances form a metric and obey triangle inequality), assuming we may only use edges from a given template graph.*

*Proof.* This is very similar to the proof of Theorem 5.12. However, in this case, we must remove from the template any edges directly from a node $i$ created in the reduction from Theorem 5.10 to the destination $d$, since any of these edges would necessarily be a shortest path (and therefore a highest preference path) from the node $i$ to $d$. Instead, we will add one additional node $i'$ for every $i \neq d$ and replace all paths of the form $(i \to d)$ with a path $(i \to i' \to d)$. We must also remove from the template any edges of the form $(x, j')$ for any $x \neq j$. Otherwise, a path ending in $(x \to j' \to d)$ would be at least as short as the same path ending in $(x \to j \to j' \to d)$, so we would not be able to enforce use of the new edges. Now we will define edges lengths $l$ as follows (from the perspective of a node $i$).

If $(i \to j \to d) \in \pi(i)$, then $l(i,j) = Q_{ij}$, $l(j,j') = Q_{ij}$, $l(j',d) = 1$. $l(i,i') = 2Q_{ii} + 1$. $l(i',d) = 1$. For two paths $(i \to j \to d) \in \pi(i)$ and $(i \to k \to d) \in \pi(i)$, define $l(j,k) = Q_{ij} + Q_{ik}$. Let $M_i = \max_j 3Q_{ij}$. $l(x,y) = M$ for all other edges $(x,y)$ (excluding the edges that have been removed from the template: $(j,d)$ for all $j$ and $(j,k')$ for all $j \neq k$).

As in the previous proof, the preference order is preserved. However, we must also verify that triangle inequality holds. Clearly, the length 1 edges obey this, since they are the shortest edges in the graph. Consider a length $Q_{ij}$ edge $(i,j)$. Any other path that starts at $i$ and ends at $j$ must either traverse a length $M_i$ edge into $j$ or a length $Q_{ij}$ edge into $j$, so this is the shortest route from $i$ to $j$. Consider a length $Q_{ij}$ edge $(j,j')$. A path that starts at $j$ must traverse either a length $M_i$ edge or a length $Q_{ij}$ edge, so this is also a shortest route. Consider a length $Q_{ij} + Q_{ik}$ edge $(j,k)$. Any path into or out of $j$ must traverse an edge of length $M_i$ or an edge of length $Q_{ij}$, and likewise for $k$. Therefore, a path out of $j$ and into $k$ must traverse at least $Q_{ij} + Q_{ik}$. Finally, consider any length $M_i$ edge. At least one end of the edge must be at some $x$ such that $(i \to x \to d)$ is not in $\pi(i)$, and any other edge into or out of this node will also have length $M_i$. Therefore, the lengths obey triangle inequality. □

Notice, if any edge may be used, and if the preferences are based on shortest path lengths for a metric defined for each node, then there is a trivial algorithm for finding an equilibrium: each node only follows the "direct to destination" path. Since a metric must obey triangle inequality, this path length cannot be strictly longer (cannot be less preferred) than any path including additional nodes. Theorem 5.10 together with Theorem 3.3 implies that FRACTIONAL SPP is **PPAD**-hard. Therefore, it is natural to next consider special instances that might be easier to solve. For instance, in real world internet routing, we would like to see path preferences primarily based on shortest paths. What would happen if we restrict ourselves to path preferences that echo the real world? Unfortunately, by adding appropriate edge lengths to the above reduction, we show that FRACTIONAL SPP is **PPAD**-hard even if all preferences are based only on shortest path lengths.

### 5.4.3 Approximate Fractional SPP

There are two notions of approximation for FSPP : $\epsilon$-solution is defined by Haxell and Wilfong [21] and $\epsilon$-stable solution is defined by Kintali [27]. Below we present their definitions :

**$\epsilon$-solution** (Haxell and Wilfong [21]) : An $\epsilon$-solution is defined as a set $w = \{w_v : v \neq d\}$ of assignments $w_v : \pi(v) \to [0,1]$ satisfying the following:

1. **Unity condition**: for each node $v$, $\sum_{P \in \pi(v)} w_v(P) \leq 1$

2. **Tree condition**: for each node $v$, and each path $S$ with start node $u$, $\sum_{P \in \pi(v,S)} w_v(P) \leq w_u(S) + \epsilon$.

3. For any node $v$ and path $Q$ starting at $v$, one of the following holds:

    - $\sum_{P \in \pi(v)} w_v(P) = 1$, and for each $P$ in $\pi(v)$ with $w_v(P) > 0$, $P \geq_v Q$; or
    - There exists a proper suffix $S$ of $Q$ such that $\sum_{P \in \pi(v,S)} w_v(P) = w_u(S) + \epsilon$, where $u$ is the start node of $S$, and for each $P \in \pi(v,S)$ with $w_v(P) > 0$, $P \geq_v Q$.

---

$\epsilon$-SOLUTION OF FSPP : Given an instance of FSPP, find a stable weight assignment $w$ that overfills each subpath by at most $\epsilon$.

---

Using Scarf's lemma, Haxell and Wilfong [21] proved that every instance of FSPP has an $\epsilon$-solution. We observe that their proof is a polynomial time reduction from $\epsilon$-SOLUTION OF FSPP to SCARF, thus showing **PPAD**-membership of the former. For more details we refer the reader to [21].

**$\epsilon$-stable Solution** (Kintali [27]) : An $\epsilon$-stable solution is a feasible solution such that for any node $v$ and path $Q$ starting at $v$, one of the following holds:

- $1 - \epsilon \leq \sum_{P \in \pi(v)} w_v(P) \leq 1$, and for each $P$ in $\pi(v)$ with $w_v(P) > 0$, $P \geq_v Q$; or

- There exists a proper suffix $S$ of $Q$ such that $w_u(S) - \epsilon \leq \sum_{P \in \pi(v,S)} w_v(P) \leq w_u(S)$, where $u$ is the start node of $S$, and for each $P \in \pi(v,S)$ with $w_v(P) > 0$, $P \geq_v Q$.

---

$\epsilon$-STABLE SOLUTION OF FSPP : Given an instance of FSPP, find a weight assignment $w$ that is exactly feasible but may underfill a higher preference subpath by at most $\epsilon$.

---

Furthermore, we can define a new notion of approximate equilibrium which encompasses both of these previous definitions: say APPROXIMATE-FSPP. It is easy to verify that the reduction from Theorem 5.10 also reduces finding an $\epsilon$-approximate equilibrium in a preference game to APPROXIMATE-FSPP. Combined with the previous observations in this section, we get the following theorem.

**Theorem 5.14.** APPROXIMATE-FSPP *is* **PPAD**-*complete*.

□

## 5.5 Kernels in Digraphs

Let $D(V, A)$ be a directed graph. Let $I(v)$ denote the in-neighborhood of a vertex $v$ i.e., $I(v)$ is $v$ together with the vertices $u$ such that $(u, v) \in A$. A set $K$ of vertices is a clique if every two vertices in K are connected by at least one arc. A set of vertices is called *independent* if no two distinct vertices in it are connected by an arc. A subset of $V$ is called *dominating* if it meets $I(v)$ for every $v \in V$. A *kernel* in $D$ is an independent and dominating set of vertices. A directed triangle shows that not all digraphs have kernels.

A non-negative function $f$ on $V$ is called *fractionally dominating* if $\sum_{u \in I(v)} f(u) \geq 1$ for every vertex $v$. The function is *strongly dominating* if for all $v$, $\sum_{u \in K} f(u) \geq 1$ for some clique $K$ contained in $I(v)$. A non-negative function $f$ on $V$ is called *fractionally independent* if $\sum_{u \in K} f(u) \leq 1$ for every clique $K$. A *fractional kernel* is a function on $V$ which is *both* fractionally independent and fractionally dominating. When it is also strongly dominating, it is called a *strong* fractional kernel. As in the integral case, a directed triangle shows that not all digraphs have fractional kernels.

An arc $(u, v)$ is called *irreversible* if $(v, u)$ is not an arc of the graph. A cycle in $D$ is called *proper* if all of its arcs are irreversible. A digraph in which no clique contains a proper cycle is called *clique-acyclic*. Aharoni and Holzman [2] proved that every clique-acyclic digraph has a strong fractional kernel. We define a computational problem – STRONG KERNEL : given a clique-acyclic digraph $D(V, E)$ with largest clique of constant size, find a strong fractional kernel. For these graphs, the proof of [2] is a polynomial-time reduction from STRONG KERNEL to SCARF. Theorem 5.15 establishes **PPAD**-hardness of STRONG KERNEL.

> STRONG KERNEL: Given a clique-acyclic digraph with the largest clique of constant size, find a fractional weight assignment to the nodes that is fractionally strongly dominating and fractionally independent.

**Theorem 5.15.** DEGREE 3 PREFERENCE GAME $\leq_P$ STRONG KERNEL.

*Proof.* We are given a preference game over player set $[n]$. We construct the following digraph $D = (V, E)$. For each player $i$, we introduce vertex $\langle i, i \rangle$ and a vertex $\langle i, j \rangle$ for each $j$ in out($i$). We have an edge from $\langle i, j \rangle$ to $\langle i, k \rangle$ if $i$ prefers $j$ over $k$. For each $\langle i, j \rangle$, $j \neq i$, we also have an additional vertex $I(i, j)$ that has an edge from $\langle j, j \rangle$ and an edge into $\langle i, j \rangle$.

We now claim that the preference game has an equilibrium if and only if $D$ has a strong fractional kernel. Let the preference game have an equilibrium $w$. Consider the following function $f$ on $V$. We set $f(\langle i, j \rangle) = w_{ij}$ and $f(I(i, j)) = 1 - f(\langle j, j \rangle)$. We have two kinds of maximal cliques. One kind is the set $\{\langle i, j \rangle\}$ for a given $i$; we have $\sum_j f(\langle i, j \rangle) = \sum_j w_{ij} = 1$. The other maximal cliques are the edges $(\langle j, j \rangle, I(i, j))$ and $(I(i, j), \langle i, j \rangle)$. Since $f(I(i, j)) = 1 - f(\langle j, j \rangle)$ and $f(\langle i, j \rangle) \leq f(\langle j, j \rangle)$, it follows that $f$ is fractionally independent.

We next show that $f$ is fractionally strongly dominating. For vertex $I(i, j)$, this is immediate since $f(I(i, j)) + f(\langle j, j \rangle) = 1$. Consider a vertex $\langle i, j \rangle$. If $j$ is the least preferred player of $i$ with $w_{ij} > 0$, then the vertex $\langle i, j \rangle$ is covered by the clique consisting of $\langle i, j' \rangle$ over all $j'$ that are at least as preferred to $i$ as $j$. Otherwise, $w_{ij} = w_{jj}$, in which case $f(I(i, j)) + f(\langle i, j \rangle) = 1$. Thus, $f$ is strongly dominating.

Suppose $D$ has a strong fractional kernel $f$. We set $w_{ij} = f(\langle i, j \rangle)$. By the fractional independence property applied to the cliques formed by $I(i, j)$ and $\langle i, j \rangle$, we obtain that $w_{ij} \leq w_{jj}$.

Consider a vertex $\langle i, j \rangle$. The set of vertices with edges into $\langle i, j \rangle$ is the union of two cliques – the set of $\langle i, k \rangle$ with $k \geq_i j$, and the set $\{I(i,j), \langle i,j \rangle\}$. If $j$ is the least preferred player such that $f(\langle i,j \rangle)$ is positive, then the sum of the weights in the first clique is 1; otherwise, the sum of the weights in the second clique is 1, yielding $w_{ij} = w_{jj}$. This establishes the stability of $w$.

The graph constructed above does not satisfy the clique-acyclic property. This is because the clique formed by the set of $\langle i, k \rangle$ with $k \geq_i j$ contains proper cycles. When the outdegree of every player in the preference game is at most 3 (including the self-loop), then we can achieve the desired condition by making the following changes to the graph. Suppose the preference list of player $i$ is $i_1, i_2, i$. Then, we replace the edge $(\langle i, i_1 \rangle, \langle i, i \rangle)$ with a three-hop path $(\langle i, i_1 \rangle, J(i, i_1))$, $(J(i, i_1), K(i, i_1))$, $(K(i, i_1), \langle i, i \rangle)$. We do the same with $i_2$. Finally, we add the edges $(K(i, i_1), K(i, i_2))$ and $(K(i, i_1), K(i, i_2))$. We can verify that in any strong fractional kernel, the weight of $K(i, i_1)$ (resp., $K(i, i_2)$) would be identical to that of $\langle i, i_1 \rangle$ (resp., $\langle i, i_2 \rangle$). The remainder of the proof is same as before. The loop $(K(i, i_1), K(i, i_2))$ guarantees that no clique contains a proper cycle. $\square$

## 6 Fractional Bounded Budget Connection Game

We define a fractional variant of the Bounded Budget Connection game, as in [30, 31]. A fractional Bounded Budget Connection game (henceforth, a fractional BBC game) is specified by a tuple $\langle V, d, c, b \rangle$, and a length function $\ell_u$ for each $u \in V$, where $V$ is a set of nodes, $d \in V$ is a distinguished destination node, $c : V \times V \to \mathbb{Z}$, $b : V \to \mathbb{Z}$, and $\ell_u : V \times V \to \mathbb{Z}$ (for each $u \in V$) are functions. For any $u, v \in V$, $c(u, v)$ denotes the cost to $u$ of directly linking to $v$, and $\ell_x(u, v)$ denotes the length of the link $(u, v)$ from the perspective of $x$, if $u$ has established this link. For any node $u \in V$, $b(u)$, specifies the budget $u$ has for establishing outgoing directed links: the sum of the costs of the links established by $u$ times the amount placed on each link should not exceed $b(u)$.

A strategy for node $u$ is a weight function $w_u : V \to [0,1]$ that $u$ places on each outgoing edge $(u, v) : v \in V$ such that $\sum_{(u,v)} c(u,v) \times w_u(v) \leq b(u)$. Let $w_u$ denote a strategy chosen by node $u$ and let $W = \{w_u : u \in V\}$ denote the collection of strategies. The network formed by $W$ is simply the directed, capacitated complete graph $G(W)$, in which the capacity of the directed edge $(u, v)$ is $w_u(v)$. The utility of a node $u$ is given by $-f(u)$, where $f(u)$ is the cost of a 1-unit minimum cost flow from $u$ to $d$, according to the capacities given by $W$ and the lengths from the perspective of $u$ given by $\ell_u$. We assume that there is also always an additional edge from each node to $d$ with cost 0, capacity $\infty$, and length = some large integer $M \gg n \max_{x,u,v} \ell_x(u, v)$; we refer to $M$ as the *disconnection penalty*. In other words, if the max flow from $u$ to $v$ is $\alpha < 1$, then $f(u)$ is the cost of the minimum cost $\alpha$ flow from $u$ to $d$ plus $(1 - \alpha) \cdot M$.

> FRACTIONAL BBC: Given a set $V$ of nodes, a destination $d$, a cost function $c : V \times V \to \mathbb{Z}$, a budget function $b : V \to \mathbb{Z}$, and a length function $\ell_u : V \times V \to \mathbb{Z}$. Find a weight assignment $w_u : V \to [0,1]$ for each $u \in V$ such that (a) $\sum_{(u,v)} c(u,v) \times w_u(v) \leq b(u)$ and (b) $w_u$ minimizes the cost of a minimum cost flow from $u$ to $d$, assuming the capacity of an edge $(x, y)$ is $w_x(y)$.

**Theorem 6.1.** PREFERENCE GAME $\leq_P$ FRACTIONAL BBC

*Proof.* We use a similar reduction from a preference game to fractional BBC. Given any instance **P** of the preference game, We will create an instance **B** of fractional BBC = $\langle V, d, c, b \rangle$, where $V = S$, $d$ = an additional node, $\forall i, j \in V$: $c(i,j) = 1$, $\forall i$: $b(i) = 1$, plus length function $l_i$ for each $i \in V$, defined as follows. Let $p_i(k)$ = the number of $j$ such that $j \geq_i k$. $\forall j \neq i, l_i(j,d) = 1$, $l_i(i,j) = p_i(j)$. $\forall j \neq i, k \neq i, l_i(j,k) = l_i(k,j) = |S|+1$. $l_i(i,d) = 1 + p_i(i)$. Given a solution to **B**, define a solution to **P**: set $w_i(j)$ = the weight placed on edge $(i,j)$ (for $j \neq i$), and $w_i(i)$ = the weight placed on edge $(i,d)$.

Consider any instance **P** of the preference game, consisting of a set of players $S$ and a preference relation $\geq_i$ for each $i \in S$. We will create an instance **B** of fractional BBC $= \langle V, d, c, b \rangle$, where $V = S$, $d$ = an additional node, $\forall i, j \in V$: $c(i,j) = 1$, $\forall i$: $b(i) = 1$, plus length function $l_i$ for each $i \in V$, defined as follows. Let $p_i(k)$ = the number of $j$ such that $j \geq_i k$. $\forall j \neq i, l_i(j,d) = 1$, $l_i(i,j) = p_i(j)$. $\forall j \neq i, k \neq i, l_i(j,k) = l_i(k,j) = |S|+1$. $l_i(i,d) = 1 + p_i(i)$. Given a solution to **B**, define a solution to **P** by setting $w_i(j)$ = the weight placed on edge $(i,j)$ (for $j \neq i$), and $w_i(i)$ = the weight placed on edge $(i,d)$.

Since the total cost for all edges is 1, and the total budget for a node is 1, each node in **B** will place total weight 1 on edges adjacent to it. This exactly corresponds to the requirement that $\sum_j w_i(j) = 1$ in **P**. The possible paths for a one-unit flow from $i$ to $d$ in **B** are: (1) the path consisting of only edge $(i,d)$, which has cost $p_i(i) + 1 \leq |S| + 1$, (2) a path of the form $(i,j,d)$ through some other node $j$, which has cost $p_i(j) + 1 \leq |S| + 1$, or (3) a path including some edge $(j,k)$ for $j \neq i, k \neq i$, which has cost $> |S|+1$. Therefore, a minimum cost flow will only use paths of the form $(i,d)$ and $(i,j,d)$, so the requirement in **P** that $w_i(j) \leq w_j(j)$ corresponds to using the weight $j$ places on edge $(j,d)$ as a capacity on that edge when finding the min-cost flow. Now, we only need to show that a node's best response in **B** exactly corresponds to a lexicographically maximal weight assignment in **P**.

Suppose we have a best response for node $i$ in **B** that corresponds to a weight assignment $w$ in **P** that is not lexicographically maximal for $i$. Then, there is some assignment $w' = w'_i \cup \{w_j : j \neq i\}$ such that for some $j \in S$, $\sum_{k \geq_i j} w_i(k) < \sum_{k \geq_i j} w'_i(k)$. There must be some $k^+ \in S$ such that $k^+ \geq_i j$ and $w'_i(k^+) > w_i(k^+)$, and there must be some $k^- \in S$ such that $\neg(k^- \geq_i j)$ and $w'_i(k^-) < w_i(k^-)$. Suppose we move $\epsilon$ weight in the best response in **B** from $P_{ik^-}$ to $P_{ik^+}$. $p_i(k^-) > p_i(k^+)$, so moving this weight will decrease the cost of a minimum cost flow, contradicting the fact that this was a best response.

Suppose we have a lexicographically maximal weight assignment $w$ for **P** that does not correspond to a best response for node $i$ in **B**. Then, in **B**, $i$ could move weight from some path $P_{ij}$ to a different path $P_{ik}$ to decrease the cost of its min-cost flow. This means that $p_i(k) < p_i(j)$, or the number of nodes preferred by $i$ over $k$ is smaller than the number of nodes preferred by $i$ over $j$. Since preference relations are transitive, this implies that $k \geq_i j$. However, since $P_{ik}$ had space left, $w_i(k) < w_k(k)$, so $w$ is not lexicographically maximal. □

**Theorem 6.2.** FRACTIONAL BBC $\leq_P$ PERSONALIZED EQUILIBRIUM.

*Proof.* Consider any instance of fractional BBC. Create a player in the matrix game for each node in the BBC instance. Assign the player one action for each available edge in the BBC instance. For any hyperedge in the matrix game, a player's payoff is negative of the length of the shortest path to the destination made up of a subset of the edges represented by that hyperedge (or negative of the disconnection penalty if there is no such path to the destination). The proof that this preserves the set of equilibria is similar to the above proof for fractional SPP games. □

**Acknowledgements** : We would like to thank H. Venkateswaran for helpful discussions, Gordon

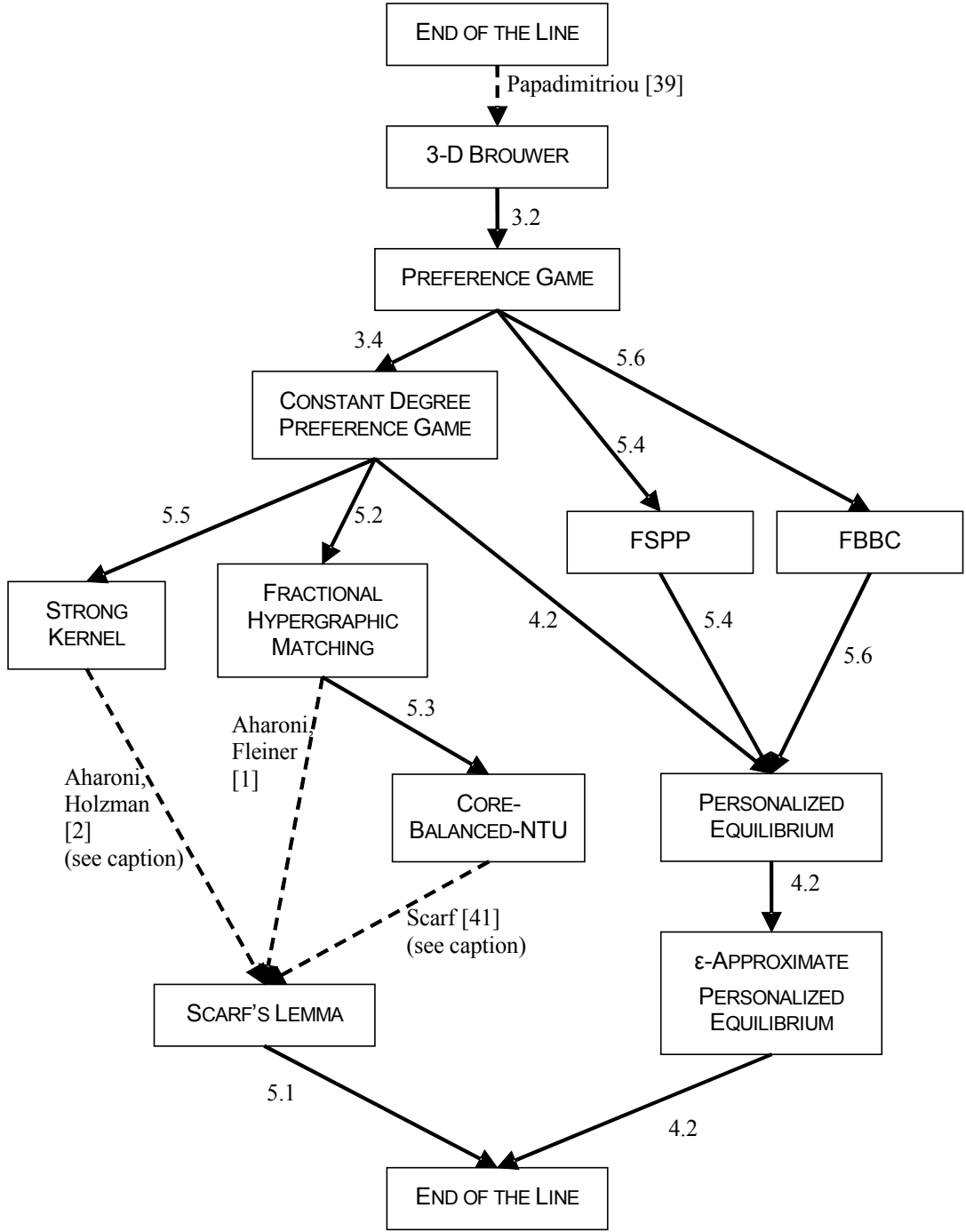

Figure 2: We show these problems to be **PPAD**-complete. Each reduction line is labeled with the Section or Citation where the reduction can be found. Two of these reductions, STRONG KERNEL $\leq_P$ SCARF and CORE-BALANCED-NTU $\leq_P$ SCARF, are only polynomial time reductions for the specific versions of the problems discussed in this paper. In our definition of STRONG KERNEL, formally given in Section 5.5, we assume that the largest clique in the graph has constant size, since otherwise it is not clear whether the problem is even in **TFNP**. CORE-BALANCED-NTU, as defined in Section 5.3, assumes that the game description explicitly lists the possible coalitions and their Pareto-optimal outcomes.